# Visualisation of financial time series by linear principal component analysis and nonlinear principal component analysis

Thesis submitted at the University of Leicester
in partial fulfilment of the requirements for
the MSc degree in Financial Mathematics and Computation

by

HAO-CHE CHEN
Department of
Mathematics University of
Leicester
September 2014

Supervisor: Professor Alexander Gorban

# Contents

# Declaration

All sentence or passages quoted in this project dissertation from other people's work have been specifically acknowledged by clear cross referencing to author, work and pages(s). I understand that failure to do this amounts to plagiarism and will be considered grounds for failure in this module and the degree examination as a whole.

Name:   HAO-CHE, CHEN

Signed:

Date:

# Abstract

In this dissertation, the main goal is visualisation of financial time series. We expect that visualisation of financial time series will be a useful auxiliary for technical analysis. Firstly, we review the technical analysis methods and test our trading rules, which are built by the essential concepts of technical analysis. Next, we compare the quality of linear principal component analysis and nonlinear principal component analysis in financial market visualisation. We compare different methods of data preprocessing for visualisation purposes. Using visualisation, we demonstrate the difference between normal and crisis time period. Thus, the visualisation of financial market can be a tool to support technical analysis.

# Introduction

The aim of this paper is to analyse various approaches to visualisation of financial time series for exploratory analysis. The result of visualisation can be found useful information to assist in making trading rules by technical analysis. We separate this paper to be three parts, (i) technical analysis review, (ii) linear and nonlinear principal component analysis comparison and (iii) data preprocessing.

By studying the review of technical analysis, technical analysis is a method to predict future prices by past prices and volumes. Financial investors can make the trading rules to make profits in stocks markets by the concepts of technical analysis. The main method of prediction the future price is to confirm the trend, because the three beliefs of technical analysis [1]: "the market discounts everything", "price moves in trends" and "History tends to repeat itself". In other words, when we find out the trend, future price can be forecasted successfully. This means that we can earn profits and avoid risks in stock market by technical analysis. On the other hand, some pervious studies ([2] and [4]) tested the resent markets. They concluded that the markets have become the efficient markets. According to the definition of the efficient market hypothesis, efficient market can be divided into three visions: weak, semi-strong and strong form of market efficiency. [3] Thus, basically the future price cannot be predicted. However, the study form Brock, Lakonishok and LeBaron [5] and the results of our three strategies testing, which are made by basic concepts of technical analysis, support the technical analysis. Taking strategy 2 for example, we success to hold the stocks in uptrend period and we avoid keeping them during the downtrend period.

We visualise the sets of financial time series (multidimensional time series). To visualise two years financial time series from 2007 to 2009 in the NASDAQ marker, we process our dataset by using log-return and principle component analysis, which can simplify the price movement relation and leaving important information when reduced dataset. And then, we compare the visualisation of two methods, linear and nonlinear principle component analysis. We can understand that the visualisation of nonlinear principle analysis represents clearer information with colouring by time series, such as clustering.

The problem of time series shifting one day is serious in our result of visualisation. We observe that the long distance move is happened widely when we shifted time series one day. So, we try to use Fourier amplitudes (or energies) to solve this problem. However, although use of the absolute values of amplitudes of discrete Fourier transform solves the problem of shifted one day successfully, a lot of clustering information is lost by transition to the absolute values of the amplitudes. This happens because we lose the information about phase shifts between different time series. Finally, we use the third method, vector of correlation coefficients of log-retunes in the moving frame, to preprocess for visualisation, and the result of visualising seems to present clustering relationship clearly, and there is no problem caused by shifting time series. It means that we consider projections of each time series onto other time series in the same frame. These projections do not change significantly in the one-day shifts and, at the same time, represent the cluster structure of the market much better because they keep the information of the phase shifts between different time series. At the end, we test this method of visualisation in the FTSE market and Taiwan stock market. The results perform as well as our testing in the NASDAQ market, which show excellent grouping information and the problem of shifted one day solved. Therefore, we conclude that the best performance of visualisation on the financial time series can be achieved through two steps. One is processing log-return dataset with the vector correlation coefficients and the other is adopting nonlinear principal component analysis to visualise.

# 1. Technical analysis review:

## 1.1 Introduction

In financial market, stock markets are a main part of high-risk investment. This means that investors may have a huge return or loss. Therefore, how to reduce the risk of loss in stock market becomes a popular task. Although some financial experts who believe efficient market hypothesis think the future price of shares is unpredictable, some traders believe that the price can be forecasted by technical analysis and fundamental analysis, such as past price and volume statistics and estimating the value of company. This part of paper is going to display the review of technical analysis, give some arguments, and apply the strategies of technical analysis in real market.

## 1.2 Technical analysis

Technical analysis is a method of prediction the trends in stock markets by past price and volume. What is more, fundamental analysis is another way to estimate the value of company. Traders who believe fundamental analysis assume that the stock price of the company should have a positive relation with the value of company. However, technical analysis traders believe security's past price and volume movements are useful in the prediction of the future prices.

Now, we focus on technical analysis. It dependents on three beliefs [1],

1. "The market discounts everything"

    Technical analysts regard that stock prices are a response to anything of company. Therefore, technical analysis can just focus on stock's price without considering about the fundamental factors of company.

2. "Price moves in trends"

    Identifying trends for each different period of time can support technical analysts to predict future price movement, because technical analysts believe the movements of the price are following the direction of trend.

3. "History tends to repeat itself"

    Chart patterns are important part in technical analysis. It helps traders to understand trends and market movements because they believe that history trends are repetitive.

Base on these three beliefs, we can claim that past prices and volumes are valuable for making trading strategies by analysing the statistics.

## 1.3 The Critics

A fruitful development of efficient market hypothesis challenges the technical analysis, while this hypothesis is to be a main criticism of technical analysis. There is a definition is given by Fama(1970, p. 383)[2]: the price of share is a reaction to all available information. Therefore, the prices of market are always the true. In other words, using any information, such as past price and chart patterns, to do technical analysis work is hopeless for efficient market, because current price just follows a random walk. ( Fama, 1970, p. 386 )

Efficient market can be classified into three visions: [3]

1. "Weak Form Of Market Efficiency"

    Past prices cannot be analysed for forecasting future prices. Current price includes all information of price during past time. And the information is a random production, so we cannot use past information and price patterns to make profit. Future price movement are followed a random walk.

**2.** "Semi-Strong Form Of Market Efficiency"

   The current price is structured by any publicly available information. Fundamental analysis is also reducing the rates of success for finding investment opportunities to produce excess returns.

   **3.** "Strong Form Of Market Efficiency"

   The all information investors can seek, no matter undercover or discover and public or private, is calculated in current price already. Thus, no investors can earn the huge profit by any information. Just the future events can affect the future price, but the probability of event happening in the future is random. Therefore, the future price's movement should be followed a random walk.

## 1.3.1 Market efficiency

Some experts provide evidences to show that markets are going to change to efficient market. This means technical analysis will become useless. Fama (1970, Pp. 388) provided some evidence by testing the weak and semi-strong form efficiency. For example, he showed that there is a table, in which nearly all values were zero for serial correlations by using natural log of price to test thirty stocks of the Dow Jones

Industrial Average from end of 1957 to September of 1962 ( Fama, 1970, Pp. 393 and 394 ). He pointed out that the relation of linear dependence among price movements were not strong. Serial correlations were a method to find the affection between repeating patterns and level of future price. In addition, he also pointed out that the profits from buy-and–hold strategy were greater than some trading rules, and this supported efficient market hypothesis. Another supporting evidence is an earlier study from Jensen and Bennington (1970)[4]. They disproved the Levy's "relative strength" trading rules, which compared the performance of target share with average performance of all stocks in market. In other words, they illustrated that "buy high and sell higher" trading rules showed a poorer performance of making profit in some conditions, and this trading rule had a higher risky than trading rules of buy-and-hold in market of New York Stock Exchange over the period from 1931 to 1965 (Jensen and Bennington, 1970, Pp. 472) [4]. According to their conclusion, the predictions of price movements were supported by the efficient market hypothesis more than Levy's trading rules on N.Y.S.E in that period of time.

Although those past studies conclude that technical trading rules are useless, technical analysis is still supported by Brock, Lakonishok and LeBaron (1992)[5]. According to their study, they claimed that it was useful to create huge returns by applying some simplest and most popular trading rules, such as moving averages and trading range break. In their data series, technical analysis was supported significantly in Dow Jones Index from 1897 to 1986. Therefore, we are going to test the trading rules in the NASDAQ Index and apply trading strategies on three stocks from 20 June 2012 to 20 June 2014. Before this experiment, it is necessary to introduce some popular and widely used concepts of technical analysis.

## 1.4  Concepts of Technical analysis

When it comes to identify price patterns, traders believe future price movements in market can be forecasted by finding the information on charts. In terms of charts, we can divide it into several concepts and we can see clear information from [1] website:

### 1.4.1  Market price trends

It is difficult to find trends from price movements, but we can define types of trend direction by high and low points, and trends can also be divided by length.

- **Uptrend, downtrend and sideways (Horizontal trend)**

*Fig 1.1. Example of uptrend*

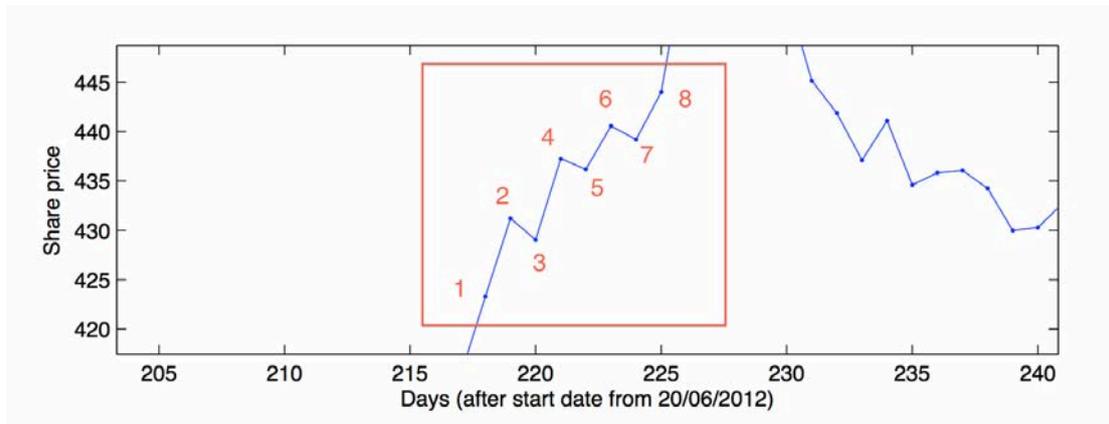

It is easy to see that peaks (2, 4, 6 and 8) and troughs (1, 3, 5 and 7) are getting higher. In this chart (Fig 1.1), the point 2 is the first high point and then price falls down to the point 3, but the point 4 is getting higher than the first high point 2 to remain an uptrend. Analogously, when peaks and troughs are getting lower, it can be called downtrend chart. However, when peaks and troughs are kept in same level of price area, we can say this is a sideways chart. (Uptrend part in [1] website)

- **Trend lengths**

We can separate trends into three terms: short-term, intermediate-term and long-term trend. Normally, short-term is less than one month, and intermediate is between one and three months. While, long-term usually lasts more than one year.

## 1.4.2 Support and Resistance

Support and resistance are another important concepts in technical analysis. We can see this to be the competition between sellers and buyers. Once the prices are going up or down closely to the line of resistance or support, it will start an opposite movements, as uptrend or downtrend. (Support and Resistance part in [1] website)

*Fig 1.2. Example of support and resistance*

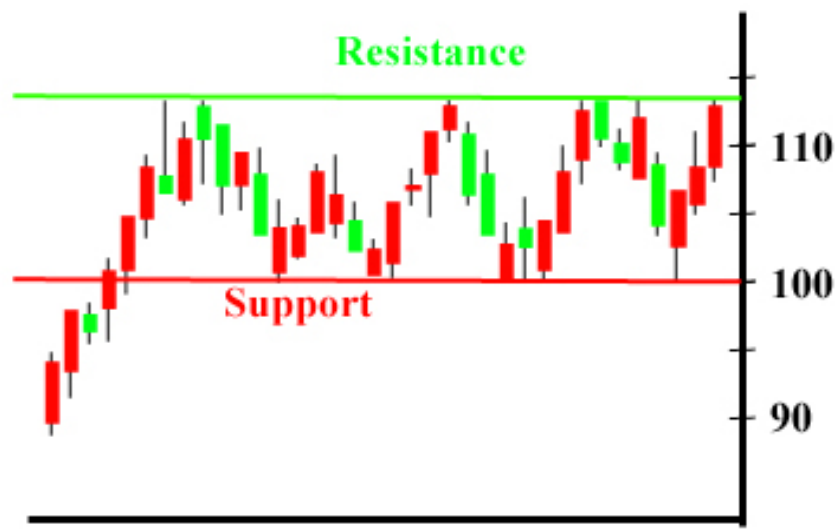

## 1.4.3 Volume

Normally, volume is the number of trading in a day. It is also an essential part of technical analysis. Traders believe that volumes are relative to future patterns, so huge number of trading is a signal for buying or selling stock. It is because that large price movement is more related to high volume in stock market. (Volume part in [1] website)

*Fig 1.3 Example of huge volume*

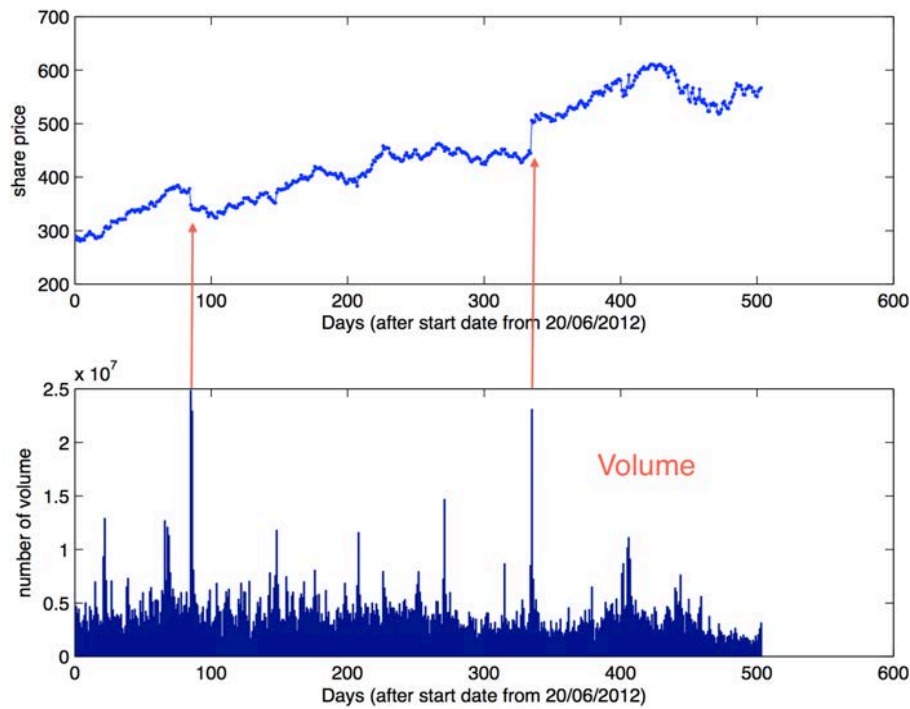

## 1.4.4  Stock chart patterns

We can find out the detail description for each chart and patterns from the [1] website.

### 1.4.4.1   Reversal and Continuation

There are two main kinds of area of patterns, reversal and continuation.

Although reversal and continuation patterns are not always trustable, they can be taken into consideration when making trading decisions.

- **Reversal:**

We can see reversal as a pressure in price structure. When buyers want to challenge the price structure, they will have sell pressure near the reversal line price. The smaller difference of price between stock price and reversal price is, the bigger sell pressure will come.

*Fig 1.4. Example of reversal*

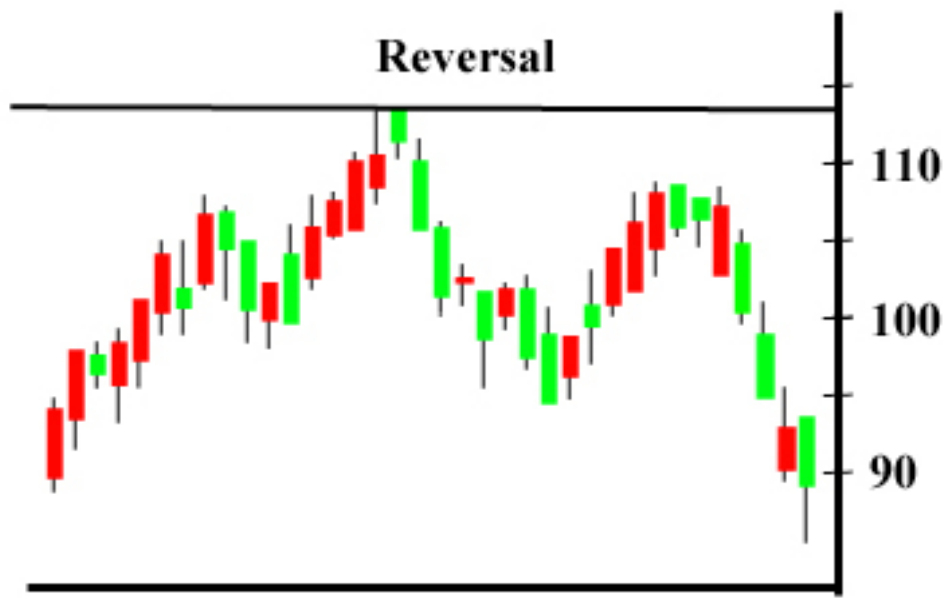

In this picture (Fig 1.4), we can clearly see that once price challenges reversal line fail,

the price will decline and the buyers start to suffer losses after that, and the trend will be a downtrend.

- **Continuation:**

Following the reversal pattern, when buyers and sellers continue to increase the pressure on the both sides, the chart will become a continuation. Normally, continuation can be classified three types, triangles (Fig 1.5), flags (Fig 1.6) and wedges (Fig 1.7).

*Fig 1.5. Example of triangle*

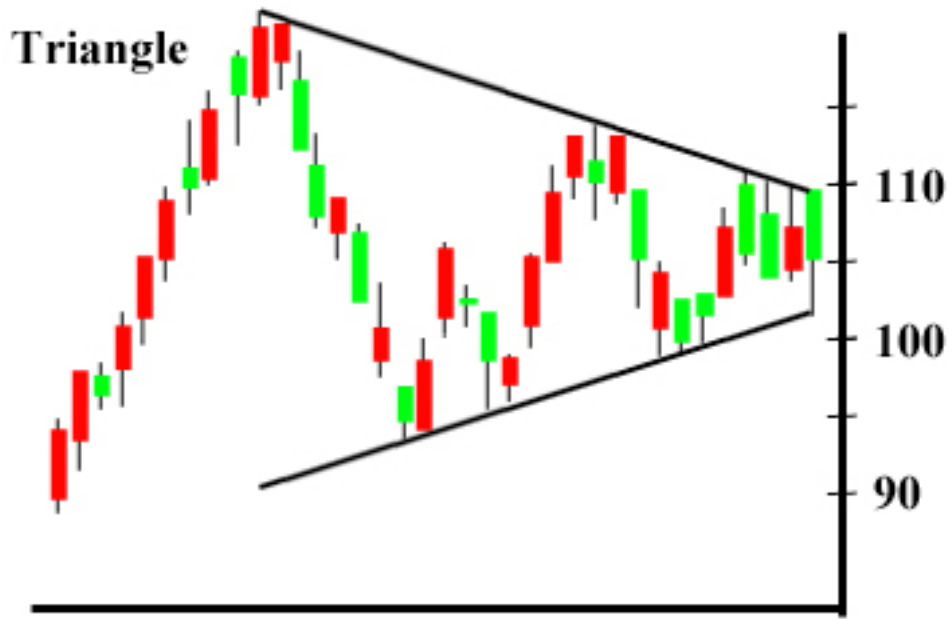

*Fig 1.6. Example of Flag*

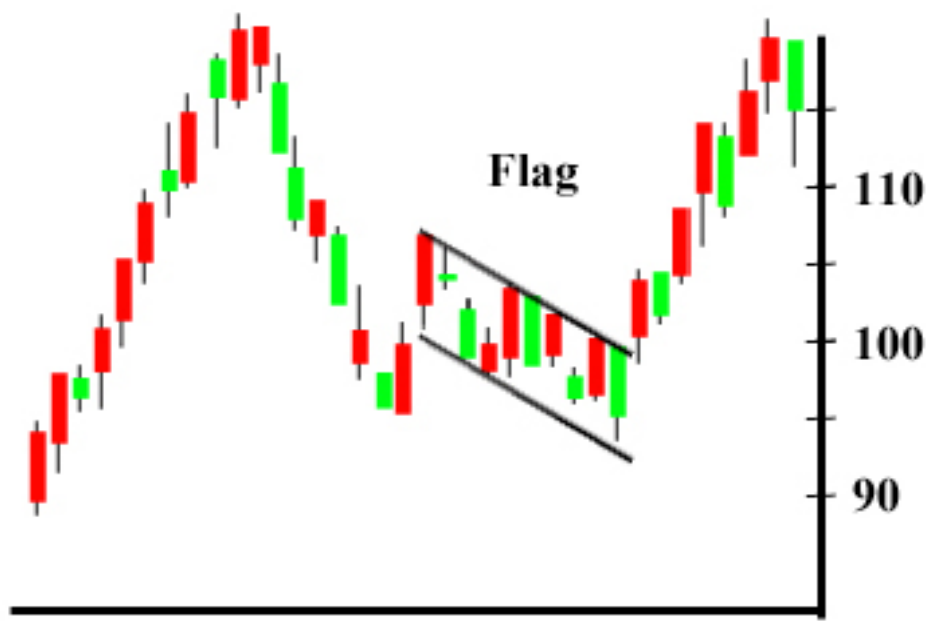

*Fig 1.7. Example of wedge*

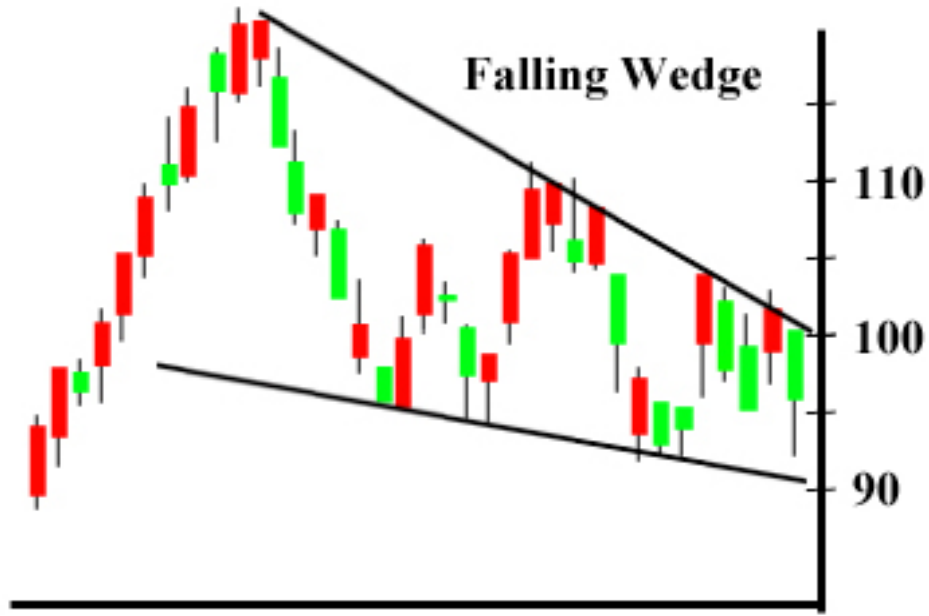

### 1.4.4.2 Head and shoulders

Here is a famous chart in technical analysis, head and shoulders chart (Fig 1.8). This structure shows the weak side of bear traders or bull traders. As shown in the below picture (Fig. 1.8), obviously, head and shoulders means trend of this stock will become downtrend. On the other hand, inverse hand and shoulders structure is a signal for uptrend.

*Fig 1.8. Example of head and shoulders for downtrend*

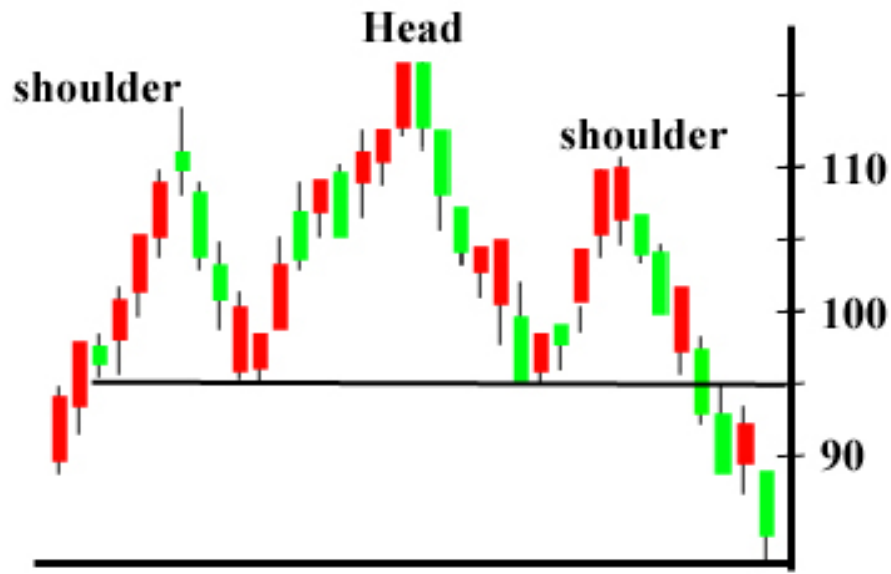

### 1.4.4.3  Cup and handle

Another widely used pattern is Cup and Handle (Fig 1.9). It is a clear signal of upward trend for traders. The shape of price pattern likes a cup with a handle. We can understand the structure like this: first part is a share price declining until a bottom and following the share price is recovered. In the end, there is a general sideways price structure.

*Fig 1.9. Example of cup and handle*

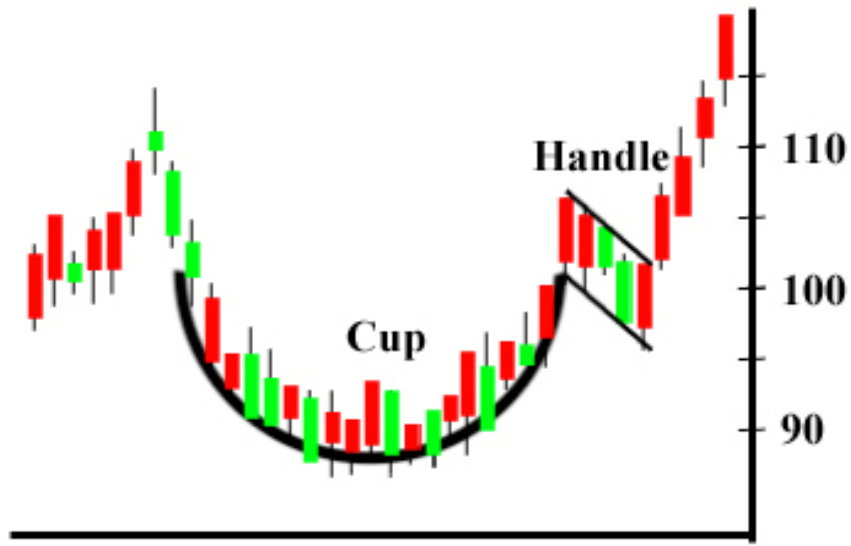

### 1.4.4.4 Rounding bottom

Rounding bottom is a chart, which likes a structure of cup and handles but without the handles. It is a long term single for changing trend form down to up.

*Fig 1.10. Example of rounding bottom*

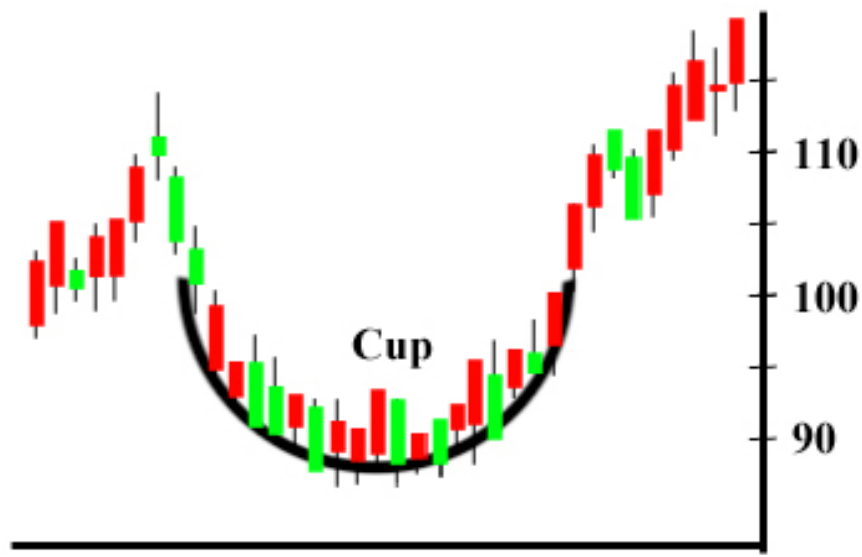

## 1.4.5 Moving Averages (MAs)

Moving averages are most popular concept for prediction of trend wards. It is a simple way to filter the noisy in real stock market by counting the past prices. In other words, it has smoothed the price patterns. Simple moving average (SMA) and exponential moving average (EMA) are two common ways to implement moving averages (MAs). [1]

- **Simple moving averages**

  Simple moving averages are the most frequently used method to count the MAs by given a number of time periods.

- **Exponential moving averages**

  Another way to calculate MAs by given weight to prices, the more closer recent date is, the significant weight it will be.

What is more, the MAs are used widely to discover the determination of trending and the levels of support and resistance. Moving Average Convergence Divergence (MACD) is an example of indicator by calculating the MAs.

*Fig 1.11. Example of moving averages*

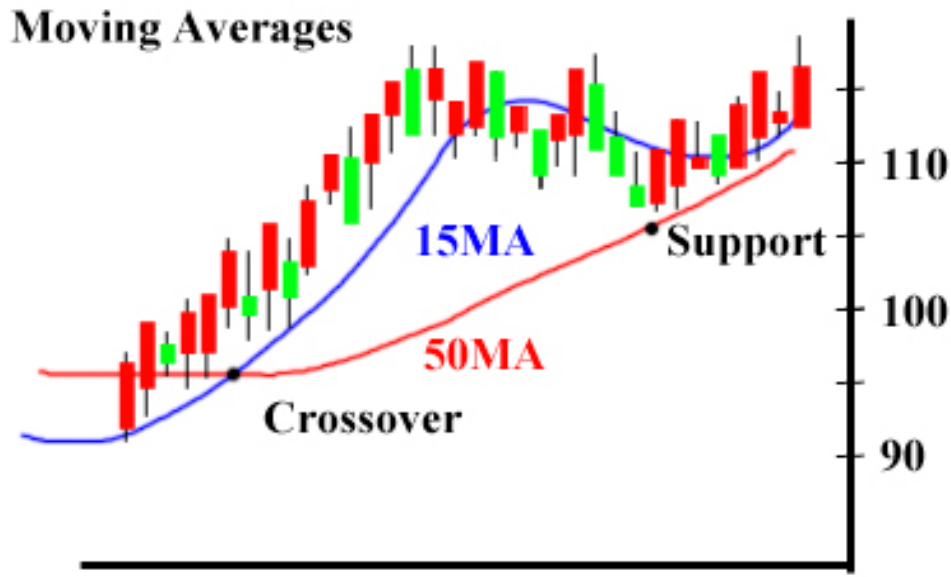

## 1.5 Strategies

After learning the concepts of technical analysis, we can combine some sample principles and pattern charts to create our strategies, and test whether it is useful or not in real stock market without trading fee to conclude technical analysis.

In terms of data, we prepare three time series form Yahoo Finance. This three time series are three companies in the NASDAQ market: Google, which is the most famous search engine in the world, Apple which produce the most popular 3C products in the world and Amazon which is the largest electronic commerce company in the world. We chose a two years period to be our sample data from 20 Jun 2012 to 20 Jun 2014, and assume we invest £2000 to each company share during this period. Finally, profits are calculated from each share for each strategy.

There are three charts for each company's closing prices and volumes:

GOOGLE:

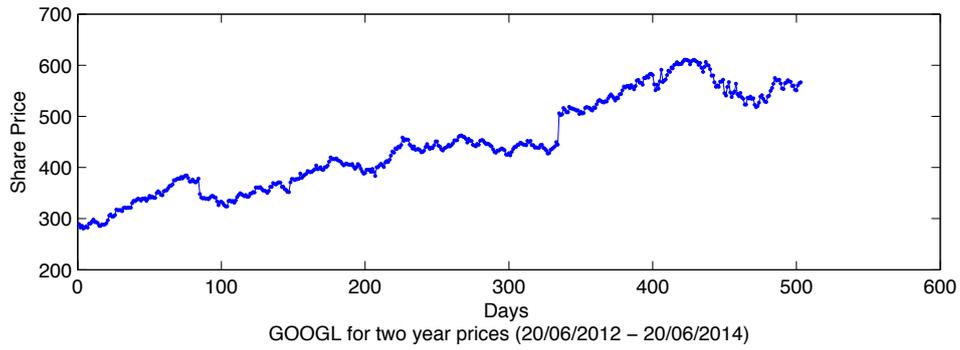
GOOGL for two year prices (20/06/2012 – 20/06/2014)

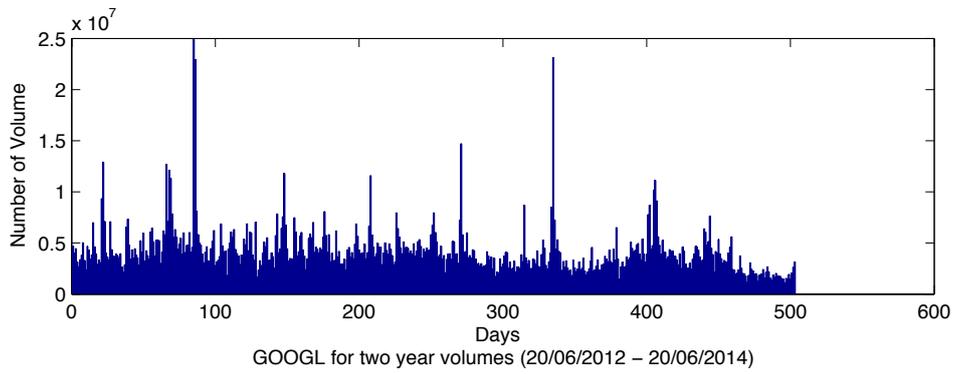
GOOGL for two year volumes (20/06/2012 – 20/06/2014)

APPLE:

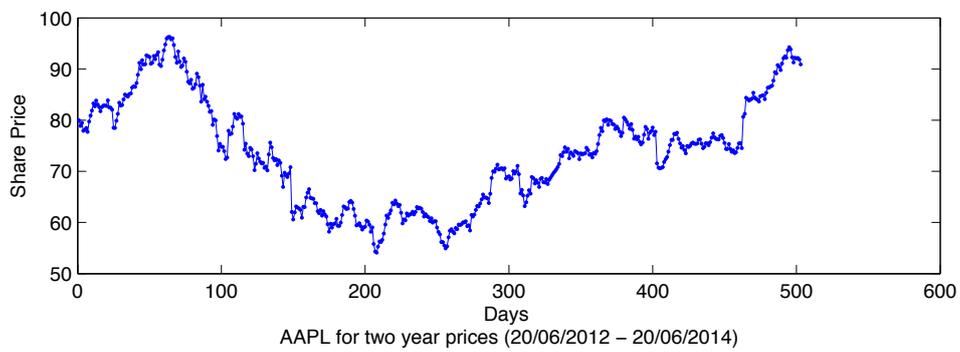
AAPL for two year prices (20/06/2012 – 20/06/2014)

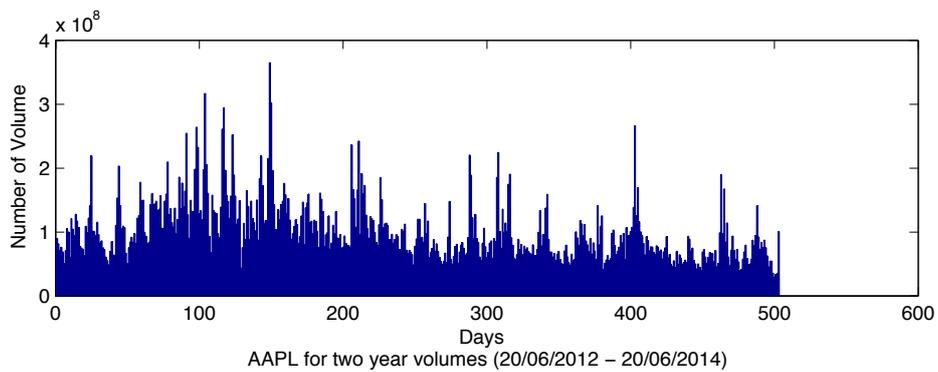
AAPL for two year volumes (20/06/2012 – 20/06/2014)

AMAZON:

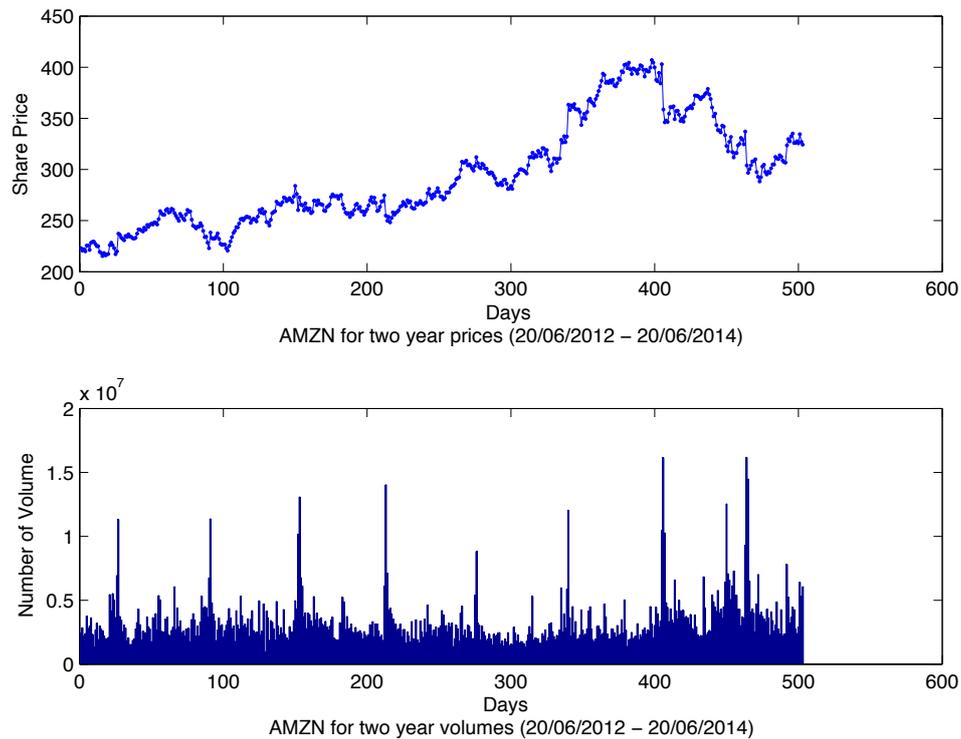

## 1.5.1 Strategy 1 - Check trend day-by-day

The idea is that we just keep the shares when they are uptrend from start day.

Otherwise, we do not keep shares. Thus, at the end of this period we should just keep the stocks, which had upward during this period. In terms of trading rules, this strategy is following the movements of price day by day, and we buy one share when price goes up 7% from base day. Otherwise, one share was sold when price goes down 7%.

We can simulate these trading recodes by running strategy 1 in Matlab.

Stock of GOOGL:

| Date | Closing-price | Action |
|---|---|---|
| '2012-07-27' | 317.8 | 'Buy' |
| '2012-08-29' | 344.35 | 'Buy' |
| '2012-09-24' | 375.07 | 'Buy' |
| '2012-10-18' | 347.85 | 'Sell' |
| '2013-01-24' | 377.48 | 'Buy' |
| '2013-03-04' | 411.16 | 'Buy' |
| '2013-05-10' | 440.56 | 'Buy' |
| '2014-04-07' | 540.63 | 'Sell' |

Stock of AAPL:

| Date | Closing-price | Action |
|---|---|---|
| '2012-08-13' | 86.39 | 'Buy' |
| '2012-08-27' | 92.66 | 'Buy' |
| '2012-10-11' | 86.13 | 'Sell' |
| '2012-11-02' | 79.1 | 'Sell' |
| '2012-11-19' | 77.93 | 'Buy' |
| '2012-12-14' | 70.23 | 'Sell' |
| '2013-01-02' | 75.63 | 'Buy' |
| '2013-01-14' | 69.12 | 'Sell' |
| '2013-02-11' | 66.5 | 'Buy' |
| '2013-02-21' | 61.81 | 'Sell' |
| '2013-04-30' | 61.35 | 'Buy' |
| '2013-06-24' | 56.15 | 'Sell' |
| '2013-07-18' | 60.22 | 'Buy' |
| '2013-08-02' | 64.51 | 'Buy' |
| '2013-08-14' | 69.99 | 'Buy' |
| '2013-09-16' | 63.2 | 'Sell' |

| Date | Closing-price | Action |
|---|---:|---|
| '2013-09-23' | 68.89 | 'Buy' |
| '2013-10-24' | 74.68 | 'Buy' |
| '2013-12-03' | 79.97 | 'Buy' |
| '2014-01-28' | 71.53 | 'Sell' |
| '2014-02-13' | 77.34 | 'Buy' |
| '2014-04-28' | 84.4 | 'Buy' |
| '2014-05-29' | 90.77 | 'Buy' |

Stock of AMZN:

| Date | Closing-price | Action |
|---|---:|---|
| '2012-08-16' | 241.55 | 'Buy' |
| '2012-09-07' | 259.14 | 'Buy' |
| '2012-10-19' | 240 | 'Sell' |
| '2012-10-25' | 222.92 | 'Sell' |
| '2012-11-23' | 239.88 | 'Buy' |
| '2012-12-18' | 260.4 | 'Buy' |
| '2013-01-25' | 283.99 | 'Buy' |
| '2013-01-29' | 260.35 | 'Sell' |
| '2013-06-10' | 281.07 | 'Buy' |
| '2013-07-12' | 307.55 | 'Buy' |
| '2013-08-16' | 284.82 | 'Sell' |
| '2013-09-18' | 312.03 | 'Buy' |
| '2013-10-25' | 363.39 | 'Buy' |
| '2013-11-29' | 393.62 | 'Buy' |
| '2014-01-31' | 358.69 | 'Sell' |
| '2014-04-04' | 323 | 'Sell' |
| '2014-04-28' | 296.58 | 'Sell' |
| '2014-06-05' | 323.57 | 'Buy' |

Result for strategy 1:

| Investing £ | Final value £ | Profit £ |
|---|---|---|
| 2000 | 2888.14 | 888.14 |
| 2000 | 2062.41 | 62.41 |
| 2000 | 2016.97 | 16.97 |

We have made **16.12%** profits by this strategy.

**Conclusion**

We can see that this simple strategy has the positive returns during this period.

However, the trading actions were traded with condition of free trade. In real market, we should be careful about times of trade.

### 1.5.2  Strategy 2 – Check trend with counters

Following the strategy 1, we want to increase our profits and avoid trading frequently. So, this time we check the trends for a period with few days for reducing the number of trades and put our money all in one trading to increase our profits. The main idea is sample way. We use two counters for checking highs and lows for determining the trend directions. And then, the values of difference between two counters are increased with movement of highs and lows. Thus, when we measure the values of difference, we can know the trend is going up or down during this period, and we believe that future price moves in trend. In addition, this time we would like to trade all of money into shares when we do buy action. On the other hand, selling all shares when we see the downtrend signal.

We can get these trading recodes by running strategy 2 Matlab codes.

Stock of GOOGL:

| Date | Closing-price | Action |
| --- | --- | --- |
| '2012-07-20' | 305.72 | 'Buy' |
| '2012-10-18' | 347.85 | 'Sell' |
| '2012-11-29' | 346.29 | 'Buy' |
| '2013-03-15' | 407.56 | 'Sell' |
| '2013-04-30' | 412.7 | 'Buy' |
| '2013-05-29' | 434.59 | 'Sell' |
| '2013-06-18' | 450.76 | 'Buy' |
| '2013-07-25' | 444.29 | 'Sell' |
| '2013-09-18' | 452.11 | 'Buy' |
| '2013-09-30' | 438.39 | 'Sell' |
| '2013-10-18' | 506.21 | 'Buy' |
| '2014-03-11' | 600.6 | 'Sell' |
| '2014-05-21' | 549.7 | 'Buy' |

Stock of AAPL:

| Date | Closing-price | Action |
|---|---:|---|
| '2012-07-19' | 83.88 | 'Buy' |
| '2012-10-01' | 90.42 | 'Sell' |
| '2013-03-15' | 61.47 | 'Buy' |
| '2013-04-05' | 58.64 | 'Sell' |
| '2013-05-02' | 61.73 | 'Buy' |
| '2013-06-19' | 59 | 'Sell' |
| '2013-07-11' | 59.6 | 'Buy' |
| '2013-09-16' | 63.2 | 'Sell' |
| '2013-10-11' | 69.19 | 'Buy' |
| '2013-11-07' | 72.37 | 'Sell' |
| '2013-11-26' | 75.33 | 'Buy' |
| '2014-01-07' | 76.26 | 'Sell' |
| '2014-02-11' | 76.14 | 'Buy' |
| '2014-06-20' | 90.91 | 'Sell' |

Stock of AMZN:

| | | |
|---|---:|---|
| '2012-07-27' | 237.32 | 'Buy' |
| '2012-10-10' | 244.99 | 'Sell' |
| '2012-11-28' | 247.11 | 'Buy' |
| '2013-02-11' | 257.21 | 'Sell' |
| '2013-03-04' | 273.11 | 'Buy' |
| '2013-03-15' | 261.82 | 'Sell' |
| '2013-04-11' | 269.85 | 'Buy' |
| '2013-08-07' | 296.91 | 'Sell' |
| '2013-09-09' | 299.71 | 'Buy' |
| '2014-01-13' | 390.98 | 'Sell' |
| '2014-02-28' | 362.1 | 'Buy' |
| '2014-03-28' | 338.29 | 'Sell' |
| '2014-04-24' | 337.15 | 'Buy' |

| '2014-05-07' | 292.71 | 'Sell' |
| '2014-05-21' | 305.01 | 'Buy' |

Result of strategy 2:

| Investing | Final value | Profit |
|---|---|---|
| 2000 | 3328.76 | 1328.76 |
| 2000 | 2637.32 | 637.32 |
| 2000 | 2606 | 606 |

We have made **42.86%** profits by this strategy.

**Conclusion**

We can find out that using strategy 2 has a significant increase of profits with 42.86%. In the same time, we also reduced the times of trade from 49 to 42. However, in the closing price of GOOGLE during this time period, the price was enhanced from under 300 to near 600. In other words, the profit for this strategy was not good enough as buy-and-hold strategy.

### 1.5.3 Strategy 3 – Volume and MAs

In this strategy, we would like to use MAs and volume to determine the sell or buy actions. As we mentioned before, MAs can smooth the data and clear the noises. Therefore, it is expected to reduce frequency of trading by introducing MAs. We design the algorithm for strategy 3 with two steps. Firstly, to check number of volume in recent three days whether it is greater than six times of average volume. Next, if both closing price and value of MAs for 5 days go up and the value of MAs for 5 days cross the value of MAs for 15 days, we recognise that as buying signal. On the other hand, when the values are going down and through the value of MAs for 15 days, it is a mark for selling.

We can get these trading recodes by running strategy 3 in Matlab.

Stock of GOOGL:

| Date | Closing-price | Action |
|---|---|---|
| '2012-07-23' | 308.06 | 'Buy' |
| '2012-10-18' | 347.85 | 'Sell' |
| '2013-01-24' | 377.48 | 'Buy' |
| '2013-07-23' | 452.35 | 'Sell' |
| '2013-10-21' | 502.15 | 'Buy' |
| '2014-01-31' | 591.08 | 'Sell' |
| '2014-02-05' | 572.17 | 'Buy' |

Stock of AAPL:

| Date | Closing-price | Action |
|---|---|---|
| '2012-07-26' | 78.5 | 'Buy' |
| '2012-10-10' | 87.89 | 'Sell' |
| '2012-12-06' | 75.39 | 'Buy' |
| '2012-12-07' | 73.46 | 'Sell' |

Stock of AMZN:

| | | |
|---|---:|---|
| '2012-07-27' | 237.32 | 'Buy' |
| '2012-10-26' | 238.24 | 'Sell' |
| '2013-01-29' | 260.35 | 'Buy' |
| '2013-01-31' | 265.5 | 'Sell' |
| '2013-07-30' | 302.41 | 'Buy' |
| '2013-10-25' | 363.39 | 'Sell' |
| '2013-10-28' | 358.16 | 'Buy' |
| '2014-01-31' | 358.69 | 'Sell' |
| '2014-03-27' | 338.47 | 'Buy' |
| '2014-04-04' | 323 | 'Sell' |
| '2014-06-20' | 324.2 | 'Buy' |

Result of strategy 3:

| Investing | Final value | Profit |
|---:|---:|---:|
| 2000 | 3104.36 | 1104.36 |
| 2000 | 2176.85 | 176.85 |
| 2000 | 2370.84 | 370.84 |

We have made **27.53%** profits by this strategy.

**Conclusion**

Using huge volume to be the signals of buying and selling is an efficient method for decreasing the times of trade, from 49 times in strategy 1 to 22 in this strategy.

Although the profits are not as good as strategy 2, it is still near 30%.

## 1.5.4 Conclusion of Strategies

Although, strategy 2 made the highest profit (42%), it was trading frequently.

Comparing with strategy 1, strategy 3 earned 27% profits just with 22 times of trading.

Thus, in my point of view, the strategy 3 is the best strategy. More important thing is that using these strategies can make profits from real market (16%, 42% and 27%, while the buy and hold returns were 51%). This means that technical analysis is useful to forecast future prices, and it is beneficial for making trading rules to avoid holding downtrend stocks at the same time.

# 2. Linear and nonlinear principle component analysis for visualisation of log-return time series

## 2.1 Dataset

In second part of paper, we focus on the visualisation of the log-return time series. In terms of dataset, log-return is a simple method to exhibit the price movement for each day. The simple formula of log-return is this:

$$L_{t+1} = log(\frac{P_{t+1}}{P_t})$$

We can see a relation between price going down and up form the formula during t and t+1 time. The log-return will return back a positive number when today (t+1) price is high than the price of yesterday (t). On the other hand, we can get a negative number of log-return when price are going down and get a zero value for keeping same price. Therefore, we think log-return can represent our time series well.

Next, after using the formula to calculate closing prices, which are ten difference time series on the NASDAQ market, we have a dataset for log-returns from 3-Jan-2007 to 3-Jan-2009. We have chosen ten stocks, GOOGL, AAPL, AMZN, BBBY, CSCO, DISCA, DTV, MSFT, SBUX and YHOO as our target stocks.

## 2.2 The problem of visualisation

Visualisation plays a key role in data analysis, because it can simplify dataset and make some complicated related information more apparent. In this part of study, we can analyse the result of visualising dataset to discover some potential relations. This results in that we need a useful tool of visualisation and a clear structure of our dataset. In visualising terms, we decide to apply in ViDaExpert software, which has a function of elastic map to present details of our dataset in two and three dimension structure. In terms of dataset, although we can separate processed closing prices into log-returns vectors, there is a serious problem for displaying such many degrees of freedom in dataset. By past study of Jackson (1991, Pp33)[6], principal components analysis can be represented by a suitable matrix with much-less dimension for dataset. Therefore, we apply the method of principal components analysis in order to reduce degrees of freedom in dataset.

## 2.3 Principal components for visualisation of log-return

Before using principal component analysis (PCA), we should understand the goals of PCA. According to Abdi. and Williams (2010, Pp. 434)[7], their study presents four

main goals for PCA:

- Important information finding form dataset.

- Keeping important information when reduced dataset.

- Description simplifying for dataset.

- Analysing the structure of dataset.

The first step we achieve principal component analysis is to create a covariance matrix $X^TX$ by data table. The matrix X means our dataset which is I observations by J variables or degrees (Abdi. and Williams, 2010, Pp.433)[7]. The sign in covariance number in the covariance matrix is meaningful. For example, positive value means both dimensions increase or decrease together and zero value shows independent relation between two dimensions. Thus, after we have covariance matrix (A), we can calculate the eigenvectors (v) and eigenvalues (λ) by formula:

$A * v = \lambda * v$.

For doing principal component analysis in our case, we have to separate our log-return dataset to be a small dimension dataset. In the experiment of this study, we divided time series to be a 20 dimensions log-return dataset. For instance, the $1^{st}$, $21^{st}$ and $31^{st}$ log-returns belong to same dimension (first column). Finally, we had a matrix of dataset in which size was 250X20 for ten time series.

After processing the dataset, we computed the eigenvalues for our covariance matrix by using formula, which we mentioned before. The below picture shows the eigenvalues of our dataset (Fig 2.1):

*Fig 2.1. Eigenvalues of log return*

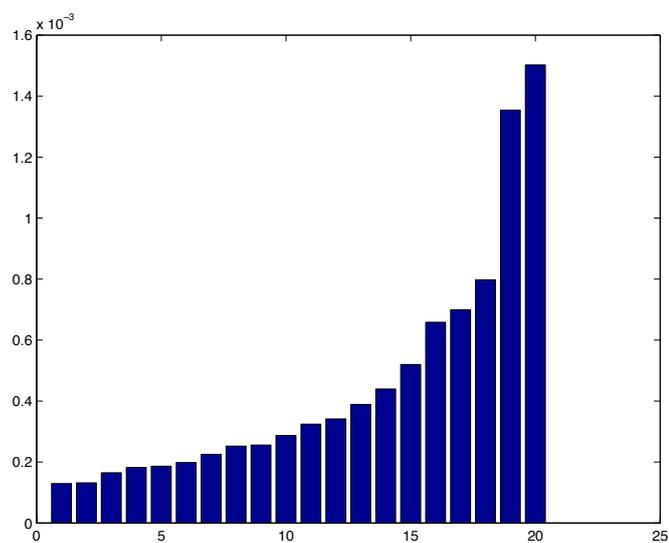

We can observe immediately that two values are larger than the others. It means that we may recover the dataset successfully by using few eigenvectors.

Moving to plotting out the log-returns, we implement the plot function by Matlab. When we plot the normalised dataset (zero-mean), we have a structure like picture a. in figure 2.2.

Following this, we chose the different group of eigenvectors to test the structure of recovered dataset with the formula of data-recovery.

Data-recovery formula:

$$Recoverd\ Data\ =\ Eignvectors \times Normalised\ Dataset$$

The idea is that we recover the datasets with different number of eigenvectors, which have the huge eigenvalue. Then, the pictures are generated to compare with picture of normalised dataset. For example, firstly, we recover the dataset with three eigenvectors (b in figure 2.2), which have biggest eigenvalues. Secondly, we made dataset with five eigenvectors (c in figure 2.2). Finally, recovering dataset with ten eigenvectors (d in figure 2.2).

*Fig 2.2. Comparing recovered dataset*

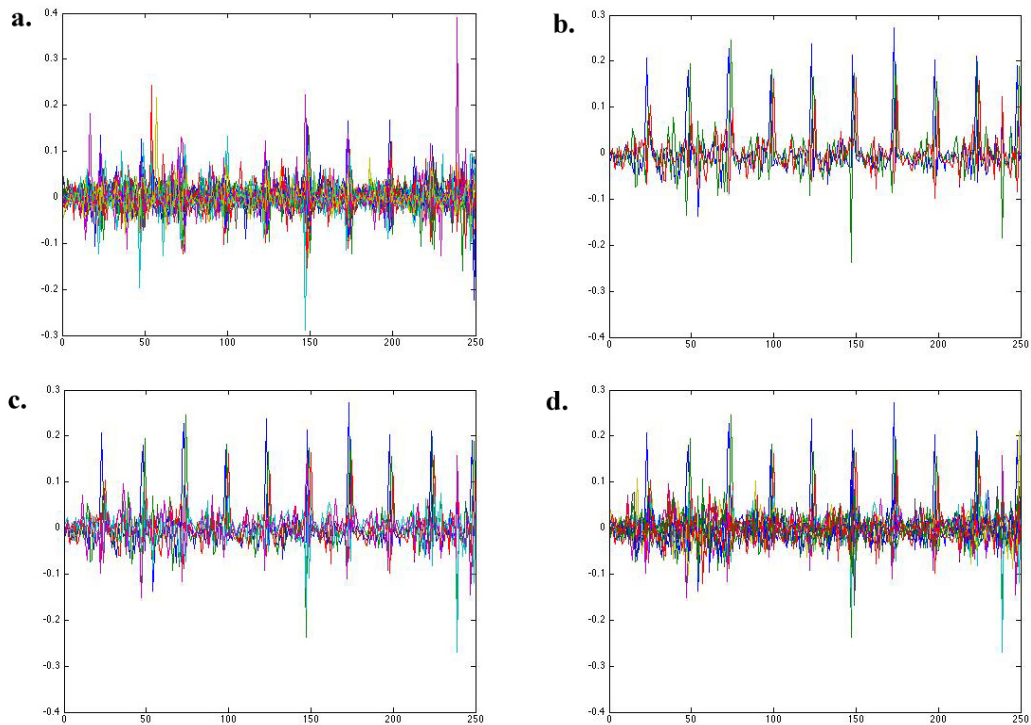

In conclusion of principal components, PCA is successful in our dataset of log-return because the pictures of structure in figure 2.2 are highly similar to the normalized dataset.

Now, we can input our log-return dataset into ViDaExpert software for displaying PCA with 3-D picture (figure 2.3). We use principal component analysis method and colour data nodes by period of times and name of stocks. In terms of structure, most of data nodes are located in the central part of three dimensions space. The outside area seems to be taken by green colour groups.

*Fig 2.3. PCA method in ViDaExpert*

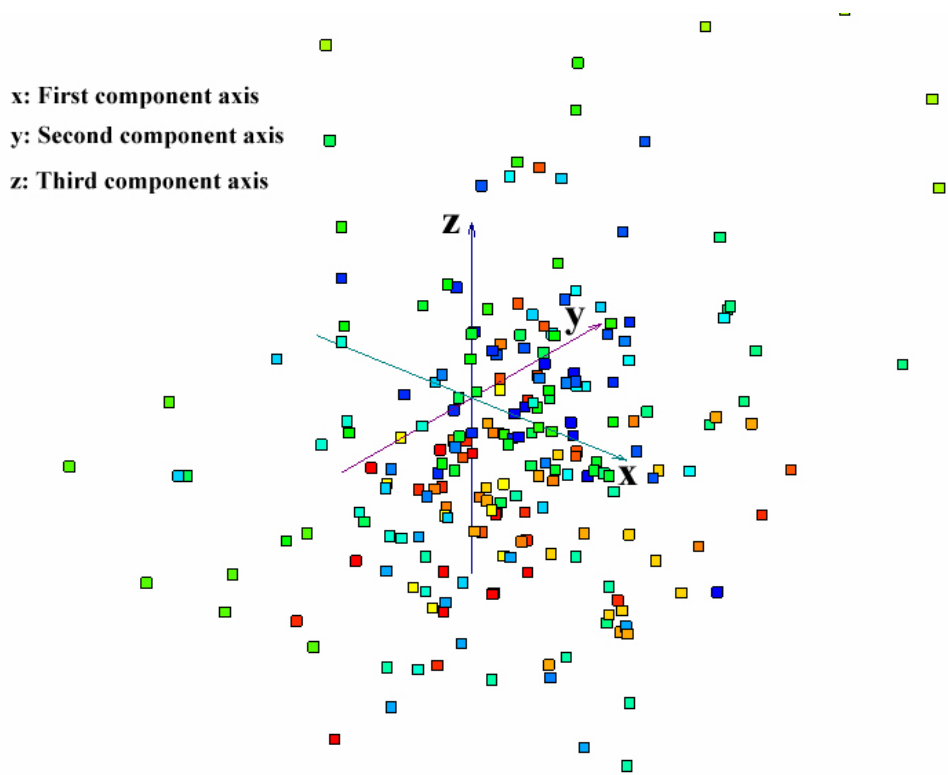

After printing out and comparing two 3-D pictures, it is clear to see that colouring by period of times shows some grouping structures on the outside area. It gives more information than colouring by stocks. Looking at figure 2.4 and 2.5, pictures show important information when we colour the data nodes in different conditions (colouring by time periods and stocks). As we can see that time periods' colouring picture (Fig 2.4) looks clearer grouping structure than colouring by name of stocks (Fig 2.5). In addition, the grouping data nodes in figure 2.4 are all at the last few periods, 23, 24 and 25, while the grouping in near central are not clear (Total 25 periods of time). Thus, we can try to discover the reason from closing price chart.

*Fig 2.4. PCA method with colouring by time periods and elastic map*

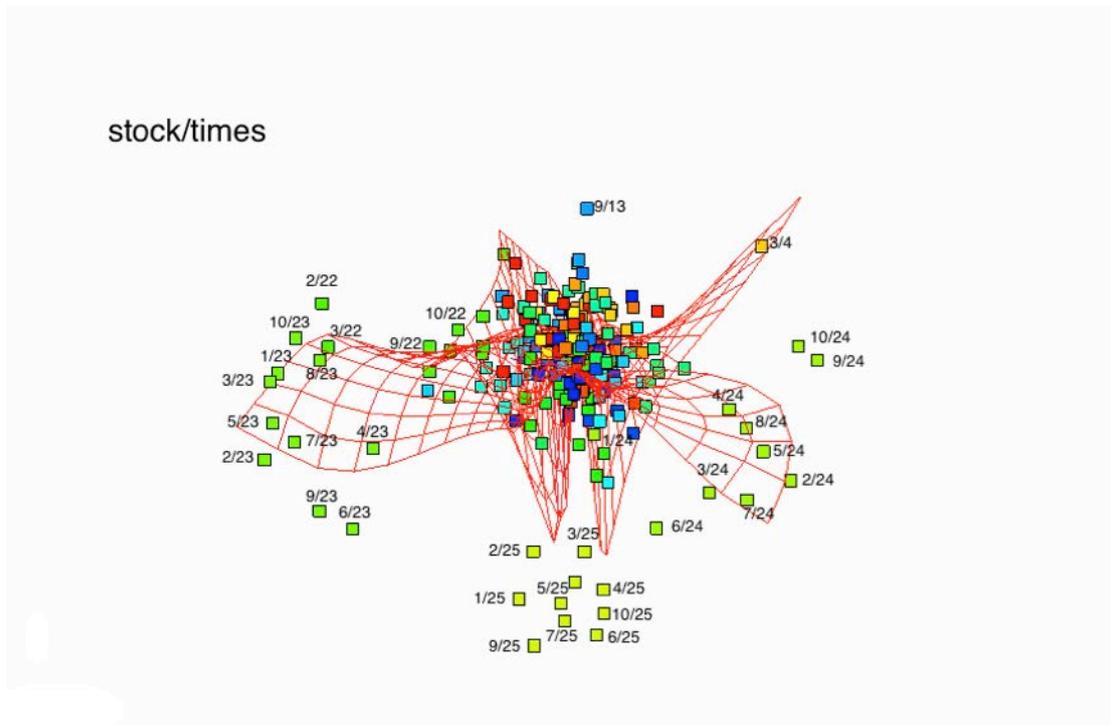

*Fig 2.5. PCA method with colouring by stocks' name and elastic map*

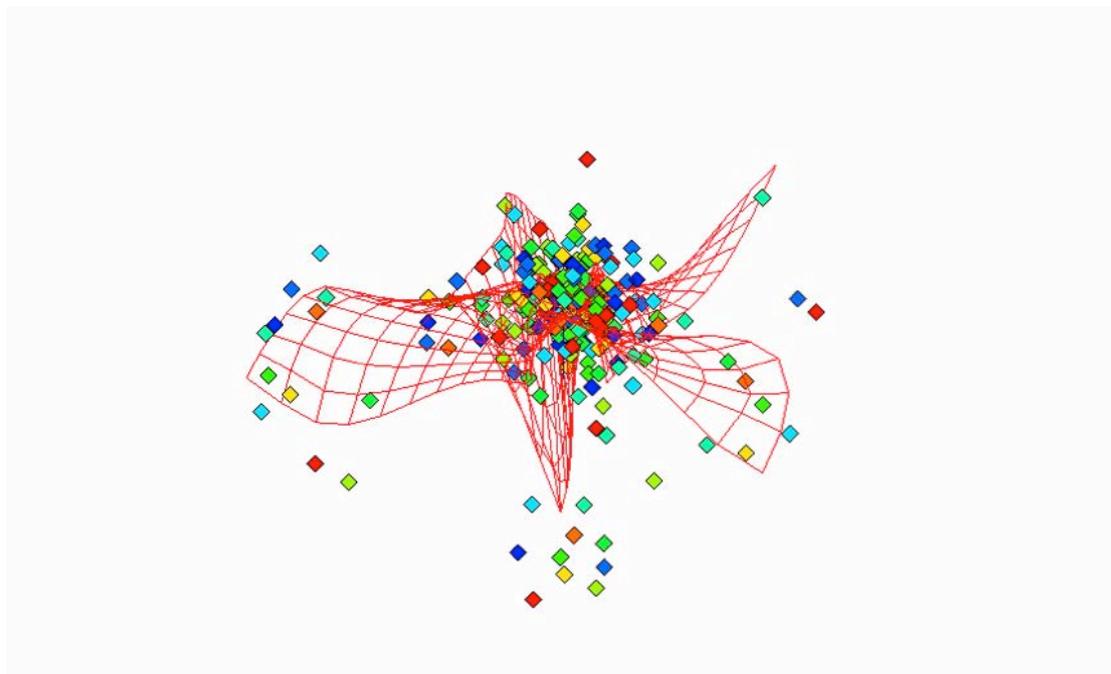

In this figure 2.6, it shows that the prices of all stock fluctuate with same trend after

the red line for each stock. This may be the reason for grouping well in outside area in

PCA of colouring time period (Fig 2.4).

*Fig 2.6. Closing price chart for marked time period at 23*

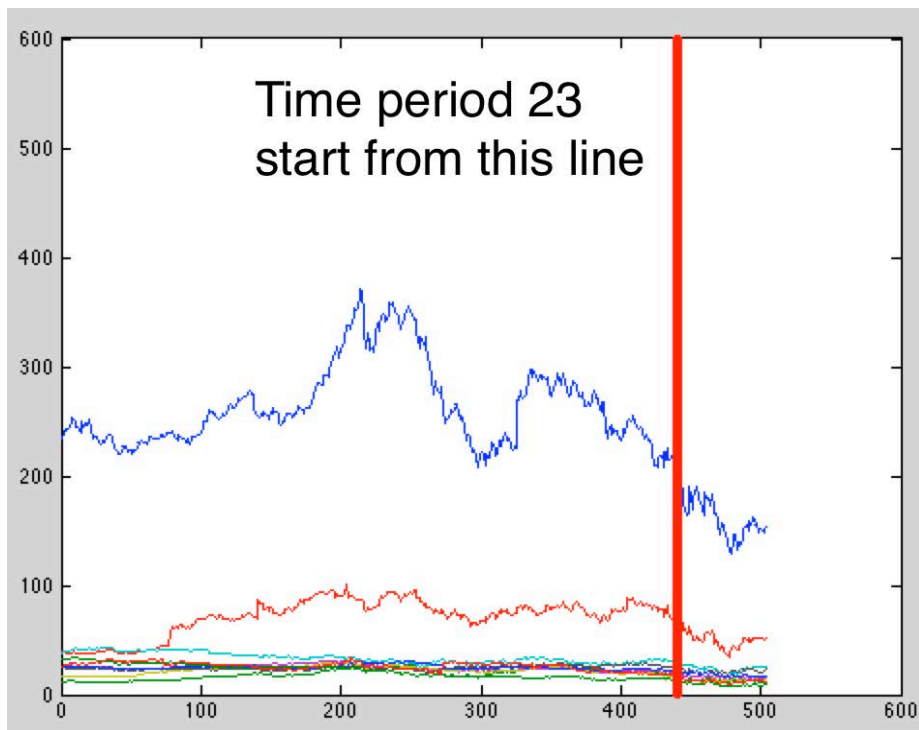

## 2.4   Nonlinear principal components methods

Nonlinear principal component analysis (NLPCA) has many methods to implement,

and we can image one of these implementations that the data nodes are projected on a plane. There is no clear structure or algorithm to achieve the NLPCA.

## 2.4.1 Kernel PCA

Taking KPCA for example, NLPCA can be derived by calculating PCA with nonlinear mapping and kernel functions: K(X,Y) = (ϕ(X), ϕ(Y)). And we also need to compute covariance matrix, but this time covariance is projected.

$$\text{Cov} = \frac{1}{N} * \sum_{I=1}^{N} \phi(X_I) * \phi(X_I)^T$$

As the process of PCA, eigenvalue is necessary, it has been rewrote like: $\lambda V = KV$, where V is eigenvectors and K is a N ∗ N matrix as a matrix with components from kernel functions (Yin, 2007, Pp.84)[8].

## 2.4.2 Elastic map

Elastic maps are contributed by a set of vertices and edges, and the key part of elastic maps for NLPCA is elastic energy functional. The general definition of an elastic map and elastic energy can be shown in example of a simple undirected graph (G) in which there are $k+1$ vertices (V) and $k$ edges (E) (Gorban and Zinovyev, 2008, Pp. 101) [9]. Once $\mu_{kj} > 0$ and $\lambda_i > 0$ are given, the elastic graph, graph (G) which is with selected families of $S_k^{(j)}$, can be embedded into a multidimensional space: $\emptyset: V \to R^m$. And elastic energy of embedding is calculated by:

$$U^\phi(G) := U_E^\emptyset(G) + U_R^\emptyset(G),$$

when $U_E^\emptyset(G) := \sum_{E^{(i)}} \| \emptyset(E^{(i)}(0)) - \emptyset(E^{(i)}(1)) \|^2$ and

$$U_R^\emptyset(G) := \sum_{S_k^{(j)}} \mu_{kj} \| \sum_{i=1}^k \emptyset(S_k^{(j)}(i)) - k\emptyset(S_k^{(j)}(0)) \|^2.$$

According to past study of Gorban and Zinovyev (2008, Pp. 103 & 104), we can consider the new locations of data nodes in multidimensional space by computing the formula of Basic Optimization Algorithm. The method of the splitting optimization likes the K-means clustering. The optimal map minimizes the energy and the K-means minimizes the distance. The definition of elastic energy for optimal map $\emptyset_{opt}$ is $U^\emptyset := U_A^\emptyset(G, X) + U(G)^\emptyset$, where $U_A^\emptyset(G, X)$ is approximation of the energy for set of data nodes (X).

Following this, data nodes are preformed as suitable distribution on grids and the map can also expand and change the level of undulating. Therefore, we can manufacture some adjustment on elastic map by coefficients μ (bending) and λ (expanding), and then seek a better projection result to observe (2008, Pp. 105). In addition, they made a clear conclusion about elastic map. Almost all liner PCA can be projected on a two-dimension elastic map. (2008, Pp.122)

## 2.5 Nonlinear principal components analysis for visualisation of log-return time series

Technically, we implement the visualisation of nonlinear principal components for our dataset by using internal coordinates function in ViDaExpert. This function can project the data points on an elastic map and we can seek more relative information from the map. In PCA part of our study, we see three clear groups in outside area when we colour data nodes by time periods.

From these pictures (fig 2.7), we discover that NLPCA method can display more grouping information than PCA method, and this is helpful to obtain more accurate clusters. However, we notice that the elastic map settings may affect the result. For example, if we make map more soft and expanded, the location of data nodes will be changed significantly (fig 2.8).

*Fig 2.7. PCA model and internal coordinate maps with colouring by time periods*

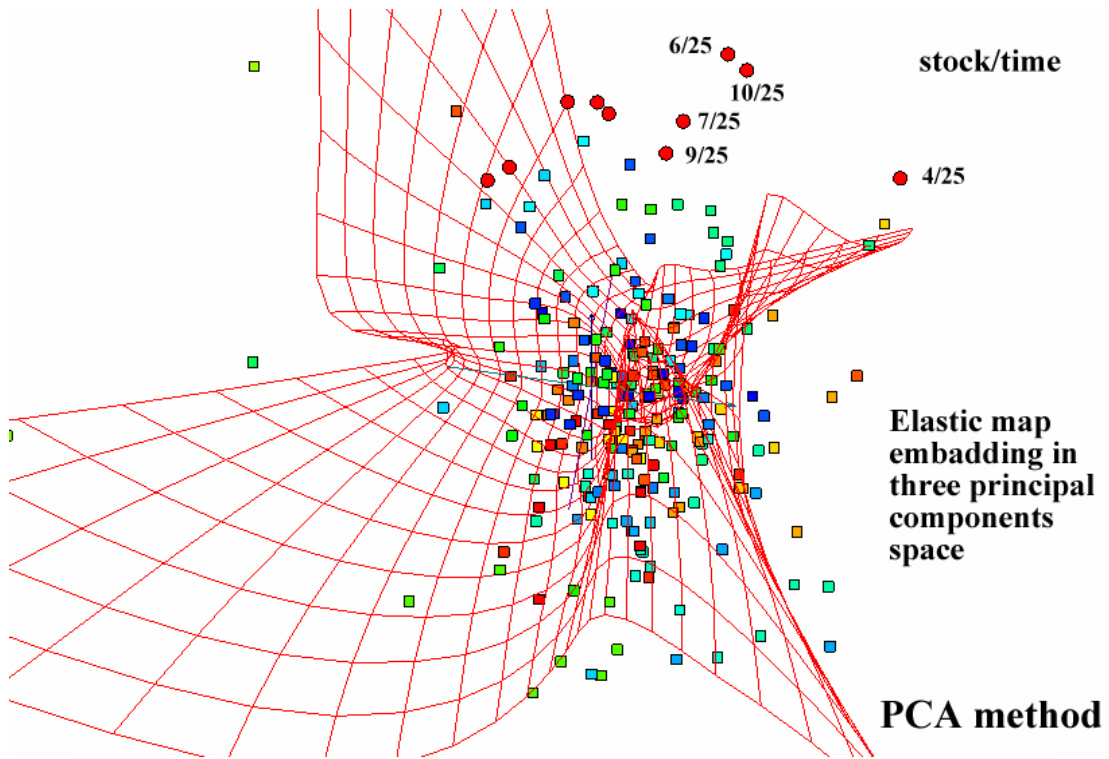

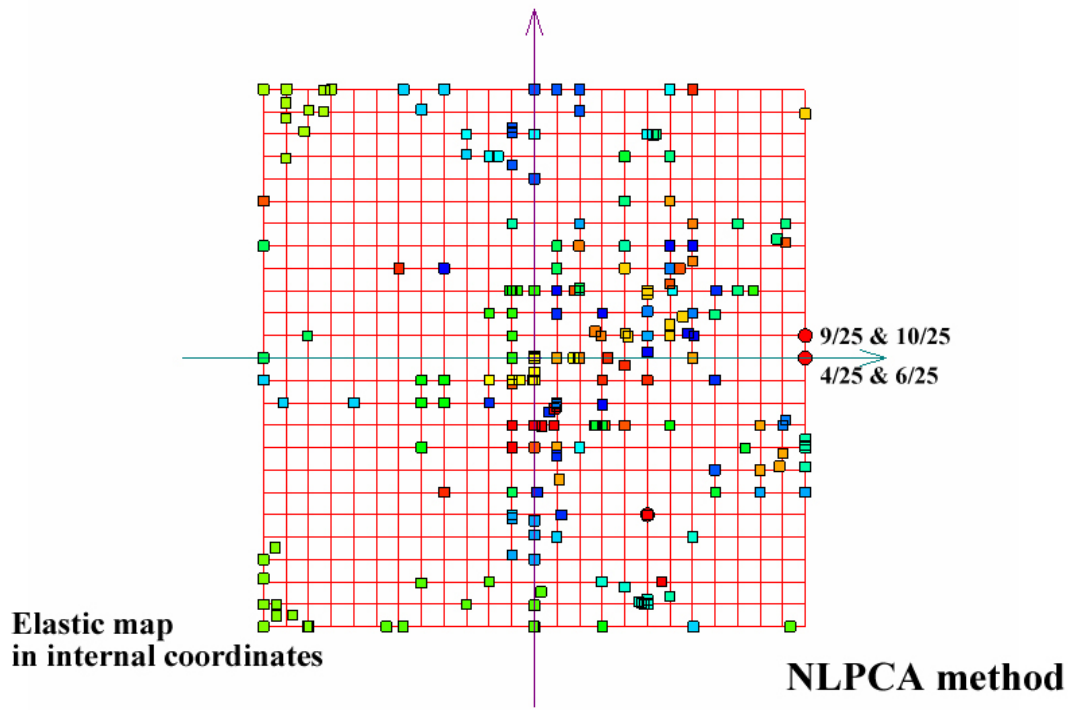

*Fig 2.8. Results of elastic map by different settings*

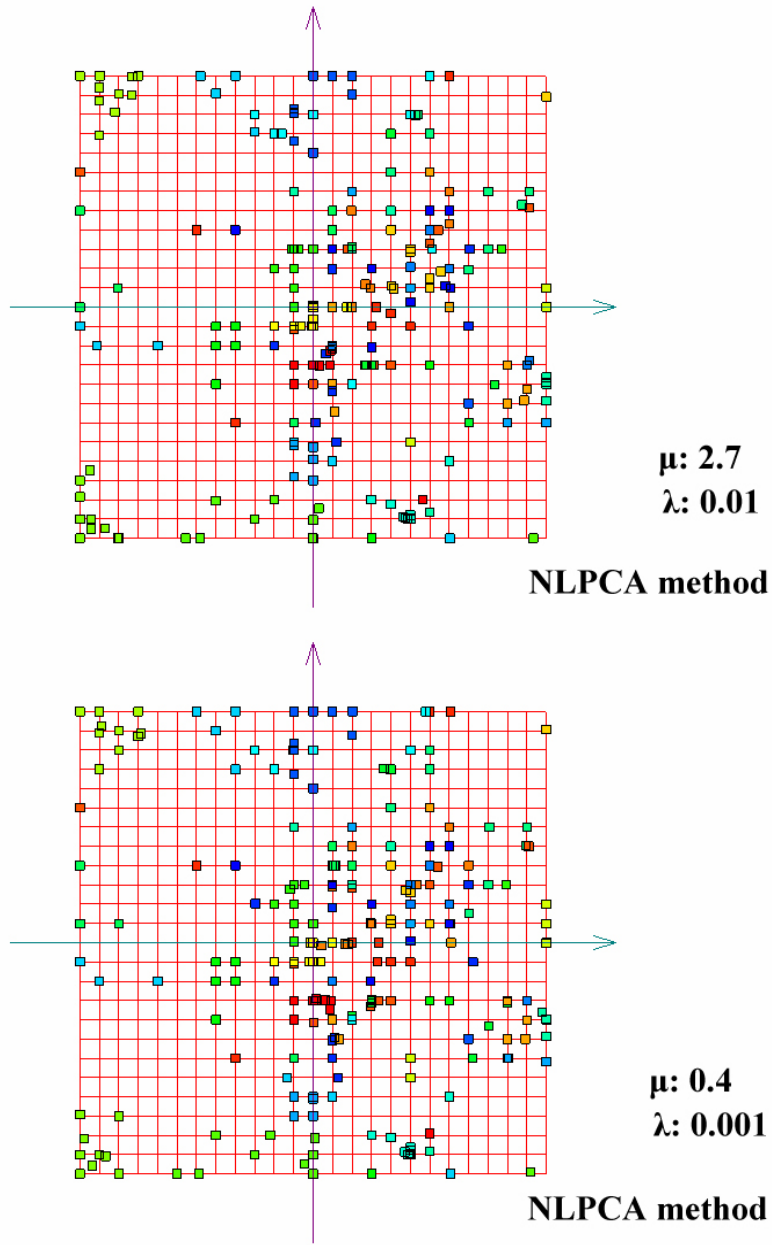

Another point of observation is that some data nodes are far from their time period group. In below picture (Fig 2.9), we have marked the data nodes in this situation. Clearly, stock 2 and 3 in many time periods have isolated locations. After reviewing closing price during this two years period (table 2.1), we discovered that only price of stock 2 and 3 increased, and prices of others dropped down more or less.

*Fig 2.9. Marking special nodes on elastic map*

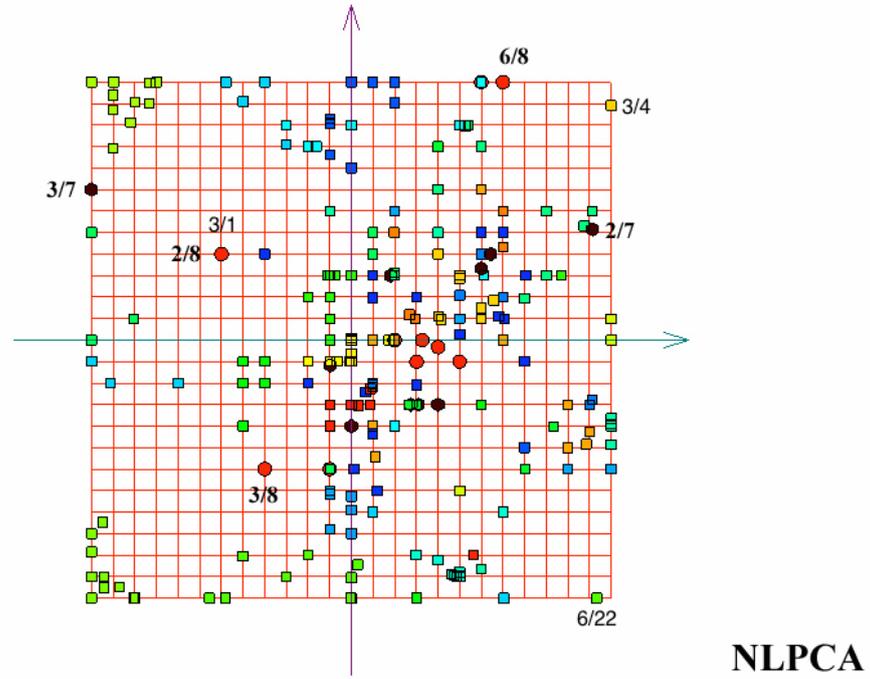

*Table 2.1. Ten stocks' closing price compare table (USD)*

| Node & Stock Date | 1 | 2 | 3 | 4 | 5 | 6 | 7 | 8 | 9 | 10 |
|---|---|---|---|---|---|---|---|---|---|---|
| | GOOGL | AAPL | AMZN | BBBY | CSCO | DISCA | DTV | MSFT | SBUX | YHOO |
| 2007-01-03 | 234.03 | 11.44 | 38.7 | 38.29 | 25.5 | 16.49 | 25 | 25.09 | 33.07 | 25.61 |
| 2008-12-31 | 153.98 | 11.65 | 51.28 | 25.42 | 14.99 | 14.16 | 22.91 | 16.86 | 8.87 | 12.2 |

## 2.6 Conclusion

A comparison of linear principal component and nonlinear principal component was presented by dataset of ten time series. Before visualising data we transformed closing price into daily log return for two years. The process of applying PCA can separate three parts: covariance, eigenvectors and eigenvalues. According to the results of visualisation from our dataset, nonlinear principal component analysis provides more detailed information than the linear method. For example, in our case, NLPCA not only shows clearly classification, but also gives information about different price movements. However, visualising NLPCA in elastic map is an unreliable method because map settings affect result dramatically and there is no unified algorithm to represent it.

# 3. Time Series Shifting for one day

To be continuing our study, we find out that there is the relation of classification among data nodes when we visualise our log-return dataset with colouring by time periods. In our expectation, the relationship of classification should be kept when we move the time series for one day. Technically, visualisation of two years dataset should not be affected significantly by shifting one day. However, the result of shifted time series is not as our expectation. Therefore, we are going to focus on time series shifting problems and represent possible solutions in this part of paper.

Based on previous results of visualisation, we conclude that more relative information is shown when we use nonlinear principal component analysis for our log-return dataset. In terms of expectation, shifting one day in time series during two years period should not effect to change the projection on elastic map dramatically. We name "jump gap" to be the difference of projection between original and shifted dataset, and we will observe the variation of jump gap on experiment.

## 3.1 Dataset

To start this experiment, it is necessary to prepare a shifted dataset from our original time series. The method of creating shifted dataset is based on the log-returns formula, which we mentioned in cheaper two, as well, and we also maintain our dataset dimensions in 20. However, for shifting one-day dataset, we have to move log-return values slightly in our vectors of new dataset. For example, second log-return values in pervious vectors are moved to first position in new vectors. Other log-return values do the same action. It also can be realised that the period of our time series is changed from 3-Jan-2007 to 4-Jan-2007 at the beginning and 3-Jan-2009 to 4-Jan-2009 at the end.

## 3.2 Problem of shifted one day in financial time series for nonlinear principal components analysis

After processed dataset of shifted one day, we use the method of NLPCA to visualise the dataset. Obviously, we can find out that the jump gap of data nodes have moved significantly between the results of original and shifted dataset at figure 3.1. To take the yellow data point 3/11 (stock number 3 and 11th time period) for example, we can see clearly that it moves the position from up and right corner in figure 3.1.a to up and left corner in b. Technically, we just have changed about ten percentages of the values in vectors from datasets. Thus, we expected that jump gaps are slighter than the phenomenon of the result.

*Fig 3.1. Comparison between NLPCA of original and shifting one-day datasets*

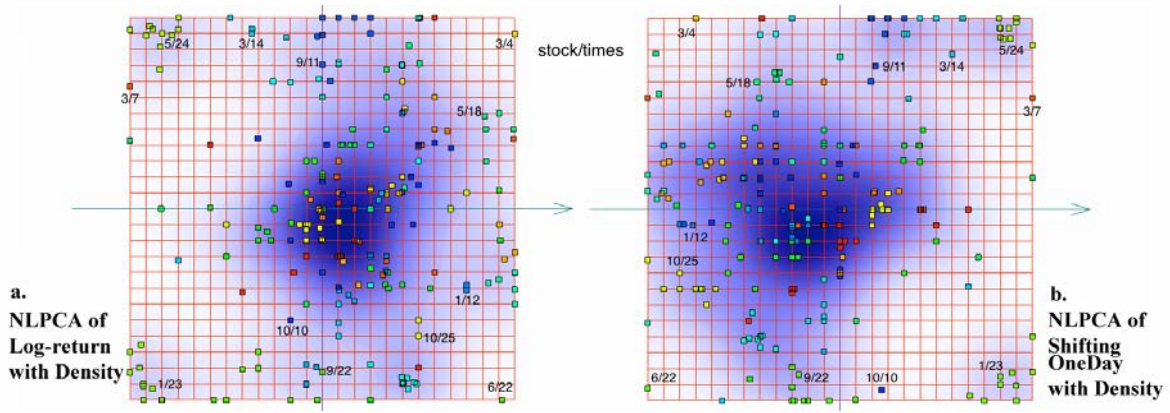

For solving the problem of significant movement in data node positions, we have to process data with a lower sensitive method for one-step time shifting. Thus, we experiment with converting dataset by the discrete Fourier transform (DFT).

## 3.3 Discrete Fourier Transform

### 3.3.1    Introduction of Discrete Fourier Transform

Mathematically, the discrete Fourier transform (DFT) is based on Fourier analysis,

and it can convert a linear function to be a periodic function during the same period [10]. Thus, we can predict that output of the DFT should not have huge difference when we shifted dataset one step. Using a simple array to be a DFT example, in figure 3.2 we can understand that Data 1 and Data 2 are similar arrays. Fundamentally, data 2 is an array transformed from data 1 by shifting one step. The most important information from this figure is that the absolute values of the amplitudes of outputs for phasing of the DFT between these two arrays are with same patterns. Therefore, based on this result, we believe that the DFT can solve the problem of the jump gap, although it may lose some variation.

*Fig 3.2 Comparison between two simple data with amplitudes of DFT*

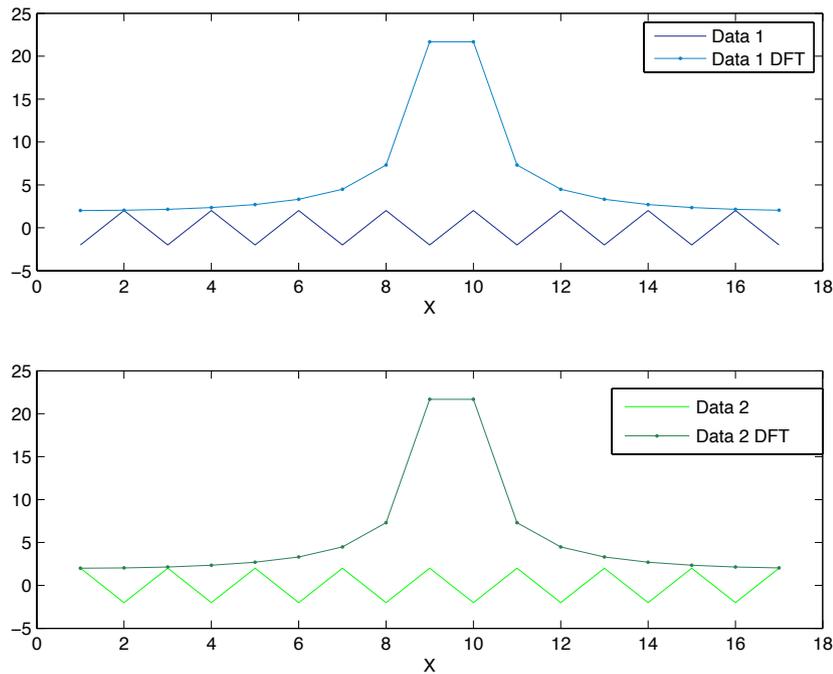

### 3.3.2 Application

We apply discrete Fourier transform (DFT) to convert our datasets by formula: [10]

$$F_k = \sum_{n=1}^{N} x_n * e^{-i2\pi kn/N}, \qquad k \in \mathbb{Z}, \qquad i = \sqrt{-1}$$

Where $x_n$ are log-returns and N is 20 for our twenty dimensions dataset.

After DFT transformation, we have the datasets with complex number format (real number + imaginary unit), which records frequency domain. Therefore, we can phase the complex numbers to be the real numbers for visualisation of NLPCA.

$$|F_k| = \sqrt{\text{real}(F_k)^2 + \text{imaginary}(F_k)^2}$$

*Fig 3.3. Comparison between NLPCA of DFT original and shifted dataset*

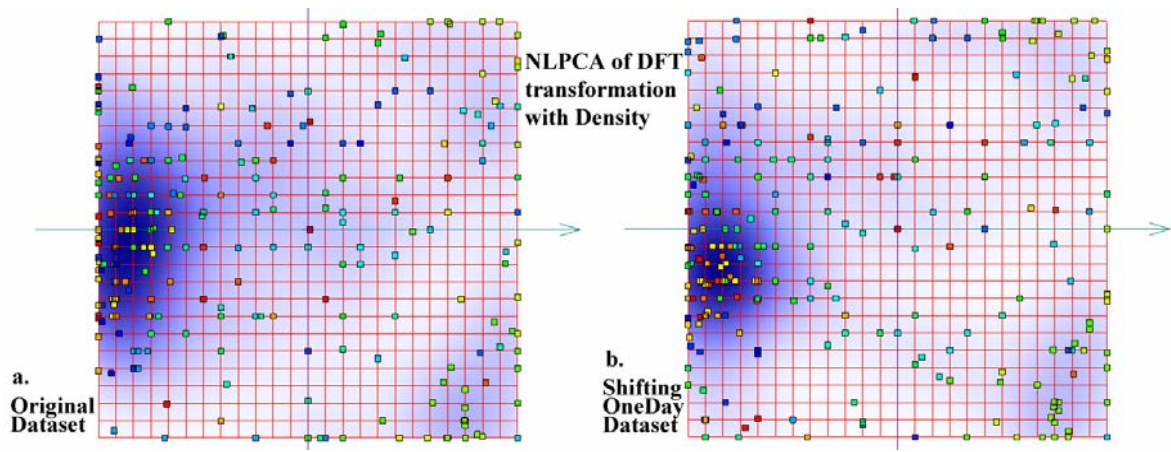

Obviously, DFT is helpful to reduce the distance of jump gaps in our datasets. In the results of the DFT visualisation (figure 3.3), most clear data node groups do not move too far. Comparing to the result on figure 3.1, it is a great success in the problem of the jump gap. What is more, we can see that clearer structures with elastic map of PCA is beneficial for us to understand the differences between unprocessed datasets and datasets with transformation of DFT (figure 3.4). It is easier to explain that the transformation of DFT is useful to decrease the level of jump gaps and maintain the structure of datasets.

*Fig 3.4. Comparison of PCA structure between un-DFT (a, b) and DFT (c, d) datasets.*
*b. and d. are one step time shifted*

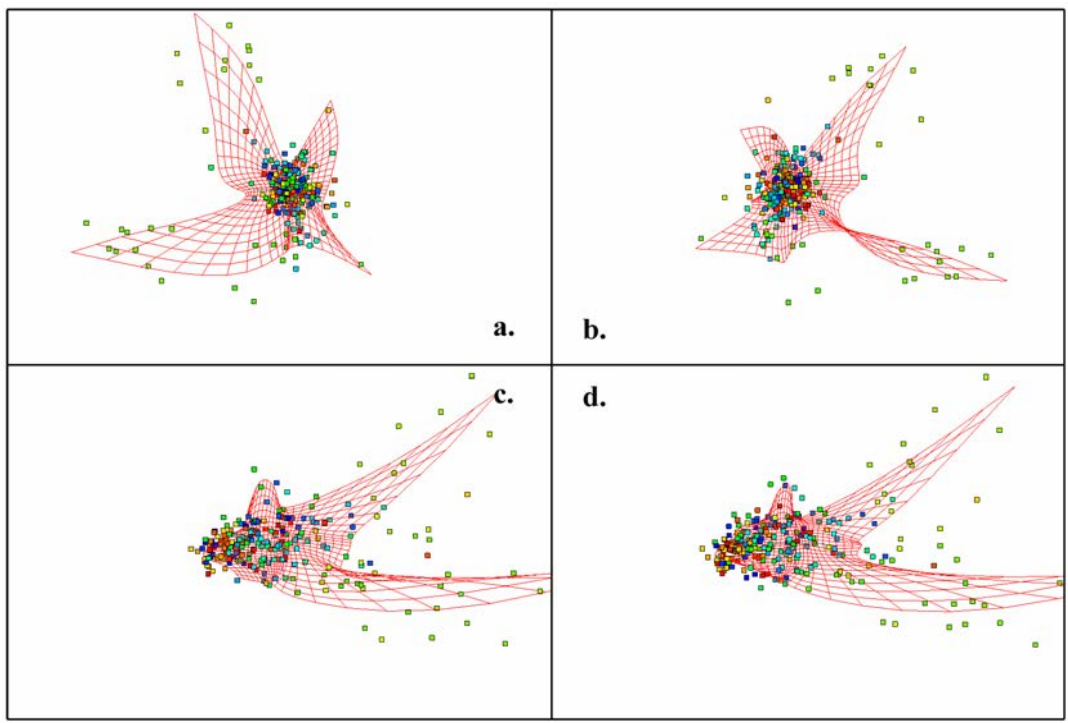

### 3.3.3 Problem of Discrete Fourier Transform

However, when we compare Fig 3.1 with Fig 3.3, we can obverse that some grouping information is lost. Clearly, there are not only projected more data points in highest density area (deepest blue colour area) but also shortened the distances between different colour groups on figure 3.3. This problem can be explained undoubtedly on the figure 3.2's example. Although the data 2 can be seen as a shifted array of the data 1, it can also be considered to be an array, which has totally different trend. This means that the different data may be classified together.

The reason is that phasing the complex number from DFT only represents the range of change. It cannot stand for price movement completely. For instance, complex numbers: 3+0.4i and -3-0.4i have totally different positions, but for computing the real numbers, | 3+0.4i | equals | -3-0.4i |. Therefore, we will divest some dissimilarities, when we use this method to calculate complex numbers in our dataset.

### 3.3.4 Conclusion

Visualisation of the log-return dataset with colouring by time period shows a better classification than by number of stocks, but when we apply the visualisation on one-day shifted dataset, it shows a significant problem of jump gaps. On the other hand, although processed log-return dataset by DFT transformation can solve the huge jumping of the data nodes, clusters are not obvious. Therefor, we want to find other methods to construct the visualisation, which has clear clustering structure and small jump for shifted dataset.

## 3.4 Vectors Projection

The projection form an n-dimensional space to an n-dimensional space is a value to represent the relation between these two vectors. So, when we consider the vectors projection, we focus on a frame (a period) not just a point (one day). According to the previous experiment of DFT transformation, we assume that problem of jump gap can be reduced by computing dataset with frame. Therefore, we will apply the vectors projection to continue our experiment.

### 3.4.1 Formula and result of vectors projection

Formulas of projection can be found in many mathematics books, we can find the basic one from website [11]. When we have two non-zero vectors, *A* and *B*. Projection *A* on *B* is calculated by

$$a_1 = |A| * \cos\theta = \frac{A * B}{|B|}$$

For applying to our dataset, we can image that one time period $(X_k^{(i)})$ is a vector with 20 dimensions (20 days' log-returns), $X^{(i)} = [X_1^{(i)}, X_2^{(i)}, ..., X_{20}^{(i)}]$. We want to record the values (*V*) for the same time period of projection on different stocks, so the formula for our dataset can be written like this:

$$V(i,j) = \sum_1^k X_k^i * X_k^j \div \sqrt{\sum_1^k (X_k^j)^2}, \quad k = 20.$$

After processed by this projection formula, we can visualise the matrix V by the method of the nonlinear principal components analysis. During the experiment, we have noted that results of visualisation are dramatically different by squaring first or summarising first in terms of division. Therefore, we represent these two results to compare figure 3.5 with figure 3.6.

*Fig 3.5. Comparison of vectors projection with time period marks, squaring first.*
*a and c are for original dataset, b and d are for shifted dataset.*

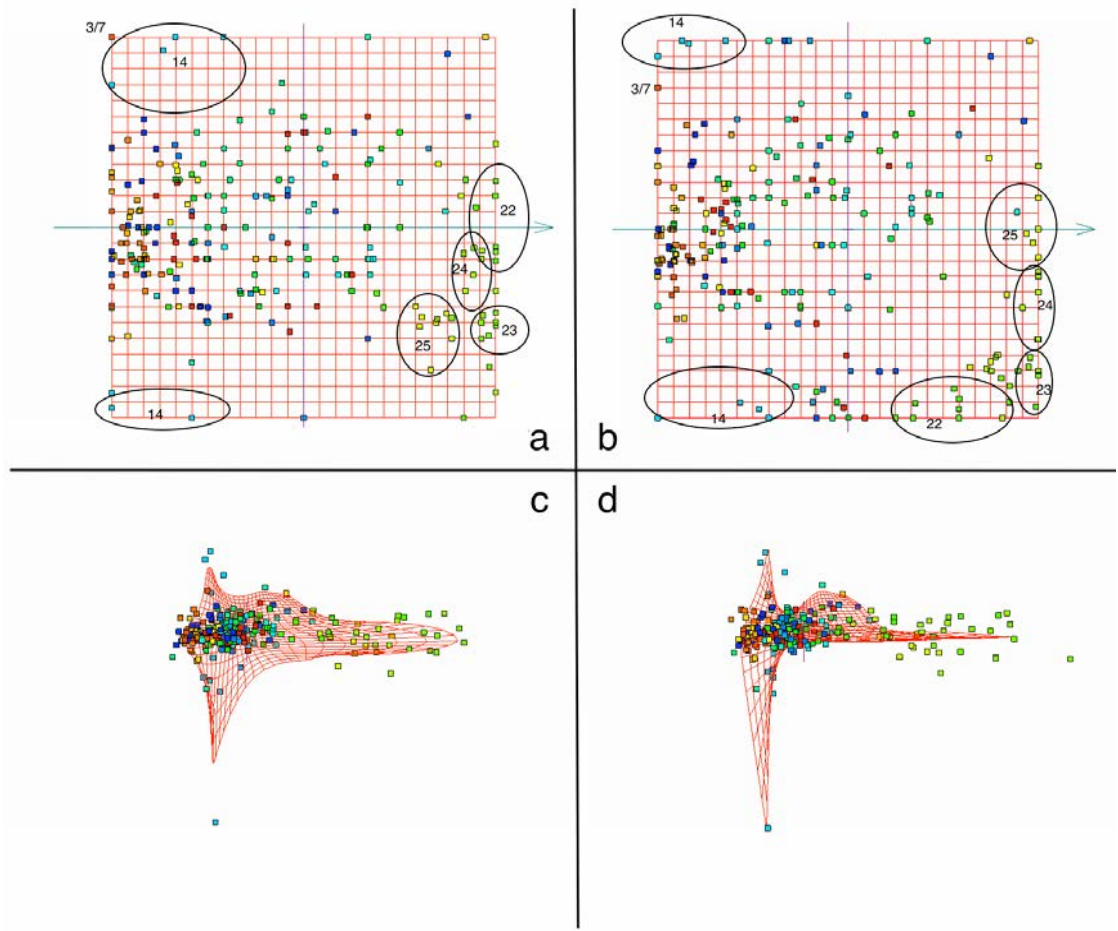

As shown in figure 3.5 a and b, the most clearest thing is that clustering is not strong. When we compare a with b, we can understand both of them have the few grouping information and the jump gaps appear not a serious problem. Comparing to previous visualisation work, we can conclude that the classification is greater than the transformation of the discrete Fourier transform and the movements of jump gaps are smaller than log-returns.

*Fig 3.6. Comparison of vectors projection with time period marks, summarising first.*

*a and c are for original dataset, b and d are for shifted dataset.*

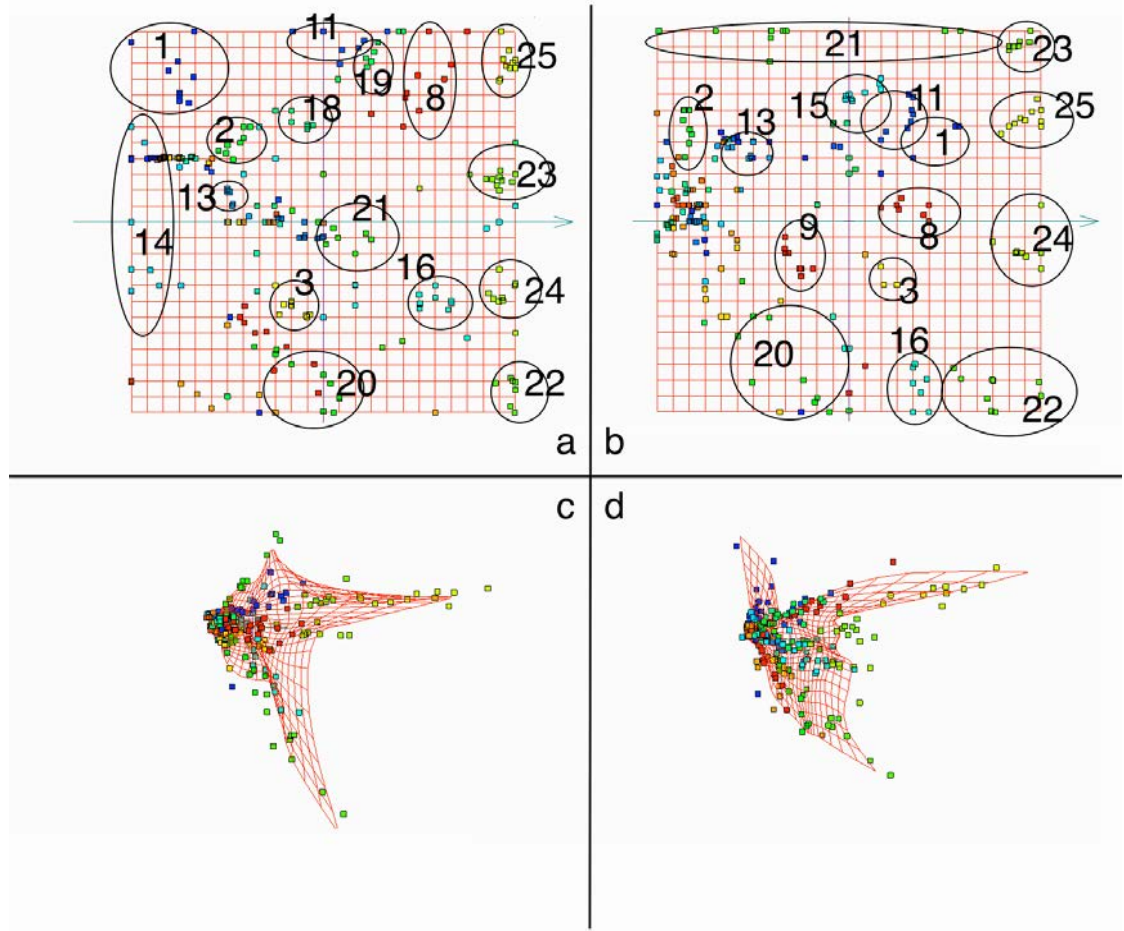

Obviously, the clustering is remarkable in this result (figure 3.6). At the same time, almost all jump gaps for time period groups are moved slightly. This method has the best performance than other pervious works. However, there is a mathematic mistake for summarizing $X_k^j$ first. We should avoid dividing the numbers, which approach zero. In our case, $\sum_1^k(X_k^j)$ could be a number less than $10^{-10}$ in some conditions. Therefore, we have to take the normalization for these special situations.

# 3.5 Normalisation of vectors projection – Vector of the correlation coefficient in the moving frame

We want to be more specific to approach the relation between the variables and avoid the problem of zero division. Pearson correlation coefficient can be considered to be our solution [12]. The formula is this:

$$corr(X^i, X^j) = \frac{\sum_k [(X_k^i - mean(X_k^i)) * (X_k^j - mean(X_k^j))]}{\sqrt{\sum_k (X_k^i - mean(X_k^i))^2} * \sqrt{\sum_k (X_k^j - mean(X_k^j))^2}}$$

The results of Pearson correlation coefficient's visualisation, which is shown in the figure 3.7, support a significant clear structure of clustering. What's more, the jump gaps are very small for each time period groups. (Compare figure 3.7 a with b) In addition, although many data nodes are collected in same area, it still can be found out some clear clustering information. In the other words, the higher density area shows the position for the normal days price movement. Therefore, to compare with others methods of visualisation, we believe that dataset processed by Pearson correlation coefficient can have best performance for visualisation.

*Fig 3.7. Comparison of Pearson correlation coefficient with time period marks.*

*a and c are for original dataset, b and d are for shifted dataset.*

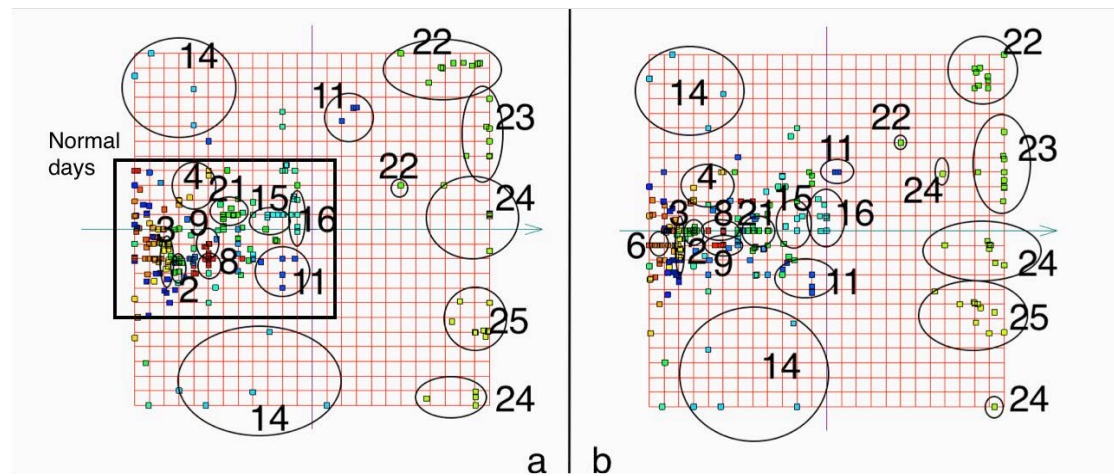

## 3.6 Apply to other stock markets

According to the result of vector projection, we can discover the visualisation showing a great clustering and slight move for one-day shifted dataset. Now, we consider visualising other stock markets by this method, and then make a conclusion on whether the method can be used in other stock markets.

In terms of markets, the dataset for pervious experiments was built from the NASDAQ, which is in the USA. We want to test this method for the markets from the different parts of the earth. Therefore, we decide to apply this method for the FTSE (the UK, Europe) and the Taiwan Stock market (Taiwan, Asia). As we choose before, we take ten time series for each market in the same time period.

*Fig 3.8. Comparison of Pearson correlation coefficient with time period marks for FTSE.*
*a and c are for original dataset, b and d are for shifted dataset.*

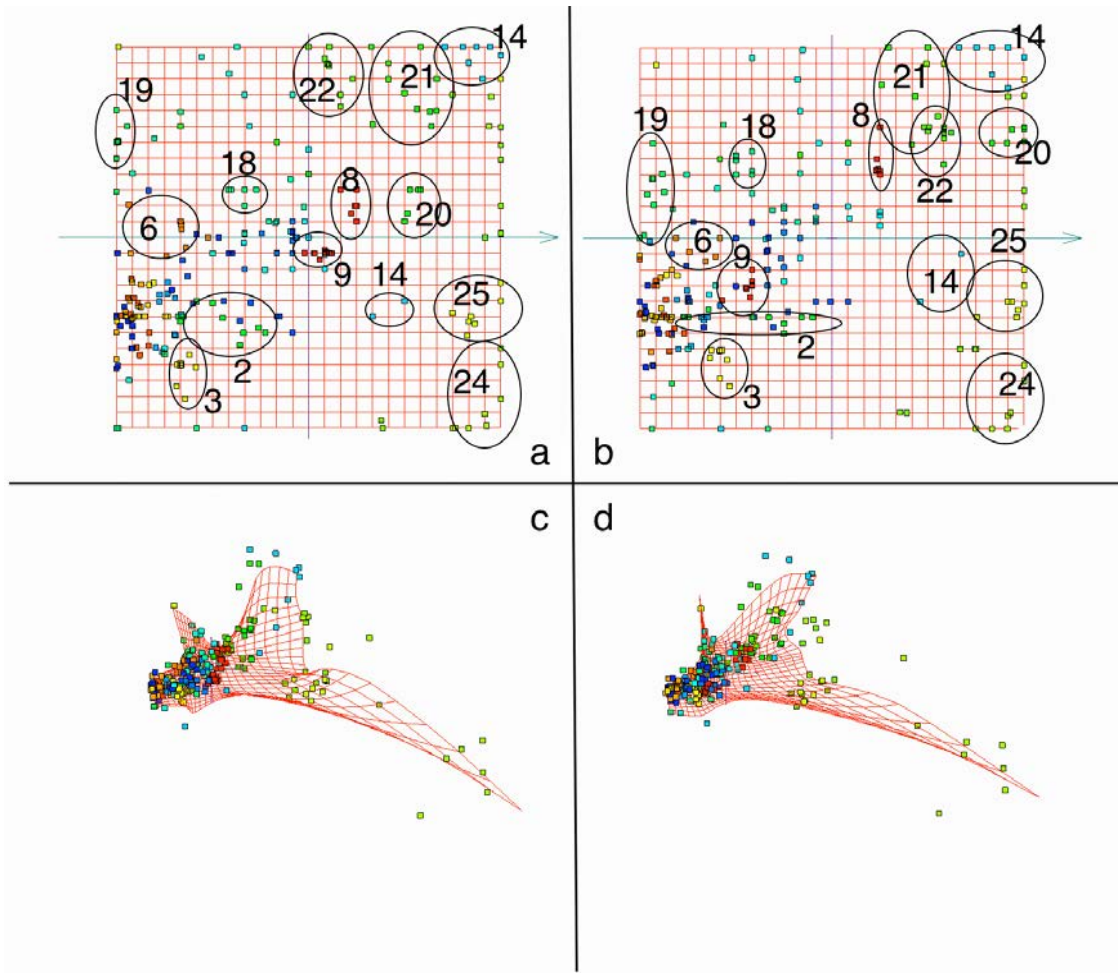

*Fig 3.9. Comparison of Pearson correlation coefficient with time period marks for Taiwan stock market. a and c are for original dataset, b and d are for shifted dataset.*

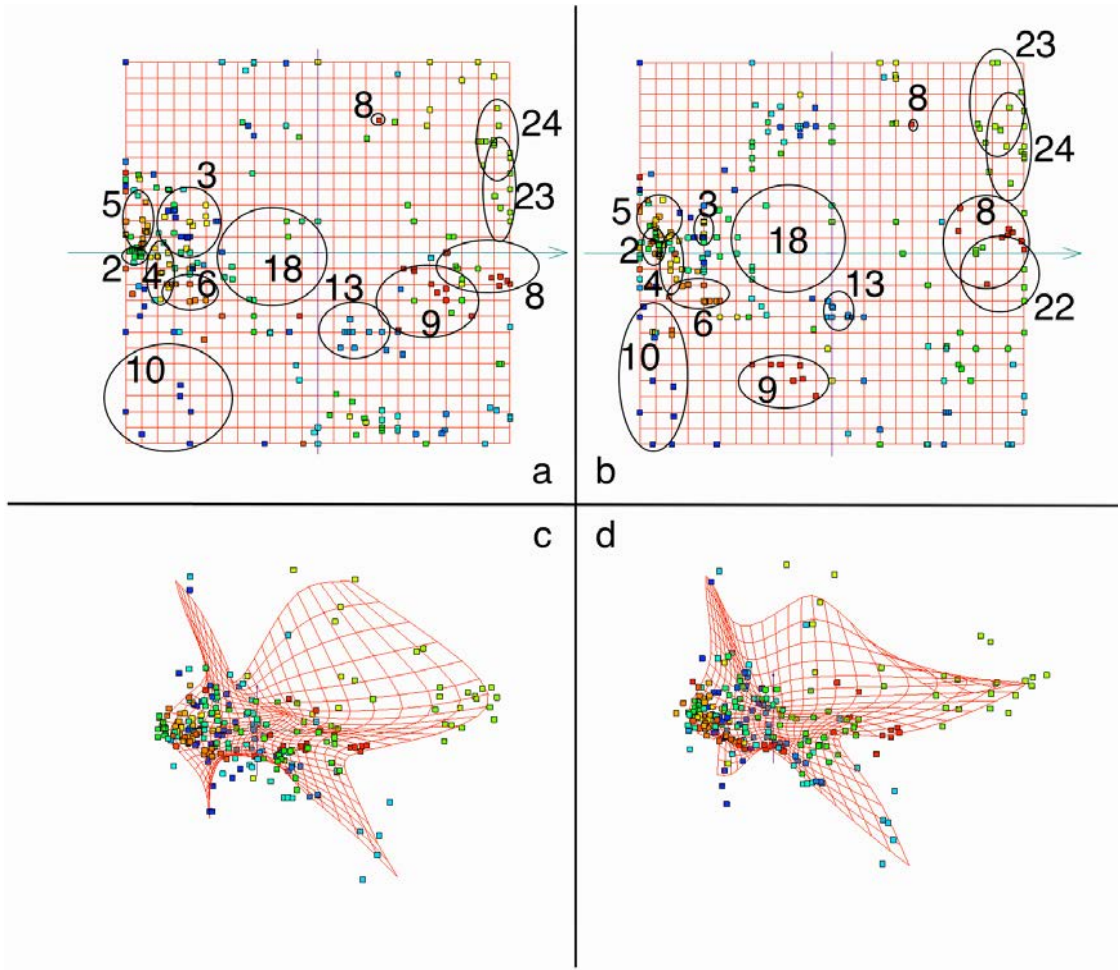

Clearly, the vectors projection can solve the problem of shifting one day in both of markets, and the visualisation for the FTSE (figure 3.8) represents a great clustering as the pervious visualisation for the NASDAQ (figure 3.7). But the results for the Taiwan stock market show an unclear grouping structure, and we can see clearly that some groups are cover each other. In addition, some data nodes are far from their grouping area. In conclusion, we can say that the time period relation in the Taiwan stock market is not as well as in the NASDAQ and the FTSE.

## 3.7 Conclusion

In this part of paper, we apply many methods to process our dataset before we visualising original and shifted dataset. At beginning, we find out the problem of shifting one day when we visualise the datasets by log-returns. Then, we try to use DFT transformation to solve the "jump gap" problem. The problem can be solved by the transformation to the absolute values of the amplitudes of DFT, but we lose much clustering information at the same time. However, we realise that we should process the dataset frame by frame to reduce the jump gaps distance. This key idea is achieved by vectors projection, and we are successful to find the Pearson correlation coefficient to normalise our dataset. Finally, the visualisation of our processed datasets displays more stable structure and clearer clusters than other methods. However, the clustering performance in the Taiwan stock market is not as clear as the NASDAQ and the FTSE.

# Conclusion

Although many stock markets seem following the description of the efficient market hypothesis, we can find some evidence to prove technical analysis is useful. We studied the basic concepts of technical analysis and made trading rules to test in the real market. As the result showing, the technical analysis helps to find a trading strategy. Thus we want to find out some useful information for building the trading rule by visualisation of time series.

In the visualising terms, after we compared the linear and nonlinear principal component analysis for visualisation of financial time series, which are processed by log-returns, we understand that nonlinear principal analysis can mine more detailed information, such as clustering. In addition, we find the jump gap problem when we shifted one day in our time series. So, we test the Fourier amplitudes (or energies) to solve the problem, and the visualisation shows a acceptable result for the problem of jump gap, but it lose a lot of clustering information. Therefore, we try to use the vector of correlation coefficients of log-returns in the moving frame, and test other markets in the same time period. According to the results, this method not only solves problems successfully, but also provides the valuable information for the trend of stock market. Thus, we believe that using nonlinear principal component analysis to visualise the dataset, which preprocesses with vector of correlation coefficients of log-returns, is a successful method for visualisation of financial time series.

# Appendix 1:
# Strategy 1.m (get time series, compute the condition and plot out)

```matlab
clear all;
clc;
target={'GOOGL','AAPL','AMZN'}; %pick up stocks
playmoney=2000;              %set money for buying stock

for i=1:size(target,2)        %get stock prices
    [mydate(:,i),a,b,myopen(:,i),myclose(:,i),myvol(:,i)]=...
        get_hist_stock_data(target(i),2012,06,20,2014,06,20);
end
w=size(myclose,2); %total number of stock
n=size(myclose,1); %total days
for i=1:w       %check miss data
    if ~isequal(mydate(:,i),mydate(:,1))
        fprintf('date not fit\n');
    end
end
%% strategy1
tradingcount=1;       %table for plot
traingtable=zeros(n,w); %counter for myrecode
for j=1:size(target,2)
    mymoney=playmoney;  %initial money
    mystock=0;          %initial number of stock
    base = 1;           %initial base
    for i=2:n
        if ((myclose(i,j)/myclose(base,j))-1)>0.07
            if mymoney>myclose(i,j)           %money enough to buy one
                mystock=mystock+1;                    %buy it
                mymoney=mymoney-myclose(i,j);         %pay it
                traingtable(i,j)=1;                   %set 1 for buy
                myrecord{tradingcount,1}=mydate{i,j};  %set trading record
                myrecord{tradingcount,2}=myclose(i,j);  %set ...
```

```matlab
                myrecord{tradingcount,3}='Buy';         %set ...
                tradingcount=tradingcount+1;
            end
            base=i;
        elseif ((myclose(i,j)/myclose(base,j))-1)<-0.07
            if mystock>0 %have enough stock to sell
                mystock=mystock-1;                  %sell stock
                mymoney=mymoney+myclose(i,j);       %take back money
                traingtable(i,j)=2;                 %set 2 for sell
                myrecord{tradingcount,1}=mydate{i,j};   %set trading record
                myrecord{tradingcount,2}=myclose(i,j);  %set ...
                myrecord{tradingcount,3}='sell';        %set ...
                tradingcount=tradingcount+1;
            end
            base=i;
        else
            continue;
        end
    end
    mymoney=mymoney+mystock*myclose(end,j);
    fprintf('finally we get: %f\n',mymoney);
    profit(j)=mymoney-playmoney;                    %compute profit
    myrecord{tradingcount,3}=target{j};             %put in record
    tradingcount=tradingcount+1;
end
PP=sum(profit)/(playmoney*size(target,2));
fprintf('we get %f percentage profit by this method\n',PP*100);

%% plot time series with trading signal
for jj=1:size(target,2)
    figure;
    
    subplot(2,1,1);
```

```matlab
        plot(myclose(:,jj),'.')

    hold on;
    plot(myclose(:,jj))

    hold on;
    for i=1:n
        if traingtable(i,jj)==1
            plot(i,myclose(i,jj),'ro')
        elseif traingtable(i,jj)==2
            plot(i,myclose(i,jj),'go')
        end
    end

str1=sprintf('Days\n%s for two year prices (20/06/2012 -
20/06/2014)\n',...
    target{jj});
    xlabel(str1);
    ylabel('Share Price');
    subplot(2,1,2);
    bar(myvol(:,jj));
str2=sprintf('Days\n%s for two year volumes (20/06/2012 -
20/06/2014)\n',...
    target{jj});
    xlabel(str2);
    ylabel('Number of Volume');
end
```

# Appendix 2:
# Strategy 2.m (get time series, compute the condition and plot out)

```
clear all;
clc;
target={'GOOGL','AAPL','AMZN'}; %pick up stocks
playmoney=2000;                 %set money for buying stock

for i=1:size(target,2)          %get stock prices
    [mydate(:,i),a,b,myopen(:,i),myclose(:,i),myvol(:,i)]=...
        get_hist_stock_data(target(i),2012,06,20,2014,06,20);
end

w=size(myclose,2); %total number of stock
n=size(myclose,1); %total days

for i=1:w          %check miss data
    if ~isequal(mydate(:,i),mydate(:,1))
        fprintf('date not fit\n');
    end
end

%% strategy2
tradingcount=1;         %table for plot
traingtable=zeros(n,w); %counter for myrecode
for j=1:size(target,2)
 mymoney=playmoney;     %initial money
 mystock=0;             %initial number of stock

 lowerP=myclose(1,j);   %initial low point
 higherP=myclose(1,j);  %initial high poin
 hcount=1;
 lcount=1;
 for i=2:n
    if myclose(i,j)>higherP   %compute the conditions for strategy2
```

```matlab
        higherP=myclose(i,j);
        hcount=1;
        lcount=lcount+1;
    elseif myclose(i,j)<lowerP
        lowerP=myclose(i,j);
        lcount=1;
        hcount=hcount+1;
    else
        hcount=hcount+1;
        lcount=lcount+1;
    end
    
    if lcount-hcount>7 %7days not go down, buy it
        %fprintf('OO trend go up at time:%d\n',i);
        higherP=myclose(i,j);
        lowerP=myclose(i,j);
        hcount=1;
        lcount=1;
        if mystock==0 %buy it if we don't have shares
            mystock=round(mymoney/myclose(i,j));   %check buy how many
            mymoney=mymoney-mystock*myclose(i,j);  %pay it
            traingtable(i,j)=1;                    %set 1 for buy
            myrecode{tradingcount,1}=mydate{i,j};  %set trading record
            myrecode{tradingcount,2}=myclose(i,j); %set ...
            myrecode{tradingcount,3}='Buy';        %set ...
            tradingcount=tradingcount+1;
        end
    elseif hcount-lcount>7 %7days not go up, sell it
        higherP=myclose(i,j);
        lowerP=myclose(i,j);
        hcount=1;
        lcount=1;
        if mystock>0 %sell it if we have shares
            mymoney=mymoney+mystock*myclose(i,j);  %take back money
            mystock=0;                             %sell stock
```

```matlab
                traingtable(i,j)=2;                    %set 2 for sell
                myrecode{tradingcount,1}=mydate{i,j};  %set trading record
                myrecode{tradingcount,2}=myclose(i,j); %set ...
                myrecode{tradingcount,3}='Sell';       %set ...
                tradingcount=tradingcount+1;
            end
        else
            continue;
        end
    end
    mymoney=mymoney+mystock*myclose(end,j);
    fprintf('finally we get: %f\n',mymoney);
    profit(j)=mymoney-playmoney;            %compute profit
    myrecode{tradingcount,3}=target{j};      %put in record
    tradingcount=tradingcount+1;
end
PP=sum(profit)/(playmoney*size(target,2));
fprintf('we get %f percentage profit by this mothod\n',PP*100);

%% plot time series with trading signal
for jj=1:size(target,2)
    figure;
    subplot(2,1,1);
    plot(myclose(:,jj),'.')
    hold on;
    plot(myclose(:,jj))
    hold on;
    for i=1:n
        if traingtable(i,jj)==1
            plot(i,myclose(i,jj),'ro')
        elseif traingtable(i,jj)==2
            plot(i,myclose(i,jj),'go')
        end
    end
```

```
str1=sprintf('Days\n%s for two year prices (20/06/2012 -
20/06/2014)\n',...
    target{jj});
    xlabel(str1);
    ylabel('Share Price');
    subplot(2,1,2);
    bar(myvol(:,jj));
str2=sprintf('Days\n%s for two year volumes (20/06/2012 -
20/06/2014)\n',...
    target{jj});
    xlabel(str2);
    ylabel('Number of Volume');
end
```

# Appendix 3:
# Strategy 3.m (get time series, compute the condition and plot out)

```
clear all;
clc;
target={'GOOGL','AAPL','AMZN'};  %pick up stocks
playmoney=2000;                   %set money for buying stock
for i=1:size(target,2)            %get stock prices
    [mydate(:,i),a,b,myopen(:,i),myclose(:,i),myvol(:,i)]=...
        get_hist_stock_data(target(i),2012,06,20,2014,06,20);
end
w=size(myclose,2); %total number of stock
n=size(myclose,1); %total days
for i=1:w          %check miss data
    if ~isequal(mydate(:,i),mydate(:,1))
        fprintf('date not fit\n');
    end
end

%% strategy 3
%moving average for 5 days
MA5=zeros(n,1);
for j=1:w
    for i=1:n
        if i>=5
            for ii=1:5
                MA5(i,j)=MA5(i,j)+myclose(i-ii+1,j);
            end
            MA5(i,j)=MA5(i,j)/5;
        else
            MA5(i,j)=0;
        end
    end
end
%moving average for 15 days
```

```matlab
MA15=zeros(n,1);
for j=1:w
    for i=1:n
        if i>=15
            for ii=1:15
                MA15(i,j)=MA15(i,j)+myclose(i-ii+1,j);
            end
            MA15(i,j)=MA15(i,j)/15;
        else
            MA15(i,j)=0;
        end
    end
end
tradingcount=1; %counter for myrecode
traingtable=zeros(n,w); %table for plot
for j=1:w
countMA=1;
mymoney=playmoney;  %initial money
mystock=0;          %initial number of stock

for i=16:n
    if MA5(i,j)>MA15(i,j)   %   5MA big than 15MA
        if myclose(i,j)>MA15(i,j) %price big than 15MA
            countMA=countMA+1;
        else
            countMA=countMA-1;
        end
    else
        countMA=countMA-1;
    end
    
    %check buy condition with big volume
    if countMA>1 & sum(myvol(i-3:i,j))>6*mean(myvol(1:i,j))
        countMA=1;
        if mystock==0 %buy it if we don't have shares
```

```matlab
                mystock=round(mymoney/myclose(i,j));  %check buy how many
                mymoney=mymoney-mystock*myclose(i,j); %pay it
                traingtable(i,j)=1;                   %set 1 for buy
                myrecode{tradingcount,1}=mydate{i,j}; %set trading record
                myrecode{tradingcount,2}=myclose(i,j);%set ...
                myrecode{tradingcount,3}='Buy';       %set ...
                tradingcount=tradingcount+1;

            end
        %check sell condition with big volume
        elseif countMA<1 & sum(myvol(i-3:i,j))>6*mean(myvol(1:i,j))
            countMA=1;
            if mystock>0 %sell it if we have shares
                mymoney=mymoney+mystock*myclose(i,j); %take back money
                mystock=0;                            %sell stock
                traingtable(i,j)=2;                   %set 2 for sell
                myrecode{tradingcount,1}=mydate{i,j}; %set trading record
                myrecode{tradingcount,2}=myclose(i,j);%set ...
                myrecode{tradingcount,3}='Sell';      %set ...
                tradingcount=tradingcount+1;

            end
        end
end

    mymoney=mymoney+mystock*myclose(end,j);
    fprintf('finally we get: %f\n',mymoney);
    profit(j)=mymoney-playmoney;                 %compute profit
    myrecode{tradingcount,3}=target{j};          %put in record
    tradingcount=tradingcount+1;

    onetrade(j)=(playmoney/myclose(1,j))*myclose(end,j); %profit table
    fprintf('onetrade= %f\n',onetrade(j));
end
```

```matlab
 PP=sum(profit)/(playmoney*size(target,2));
 fprintf('we get %f percentage profit by this mothod\n',PP*100);

%% plot time series with trading signal
for jj=1:size(target,2)
    figure;
    
    subplot(2,1,1);
    plot(myclose(:,jj),'.')
    
    
    hold on;
    plot(myclose(:,jj))
    
    
    hold on;
    for i=1:n
        if traingtable(i,jj)==1
            plot(i,myclose(i,jj),'ro')
        elseif traingtable(i,jj)==2
            plot(i,myclose(i,jj),'go')
        end
    end
    
str1=sprintf('Days\n%s for two year prices (20/06/2012 -
20/06/2014)\n',...
    target{jj});
    xlabel(str1);
    ylabel('Share Price');
    subplot(2,1,2);
    bar(myvol(:,jj));
str2=sprintf('Days\n%s for two year volumes (20/06/2012 -
20/06/2014)\n',...
    target{jj});
    xlabel(str2);
    ylabel('Number of Volume');
end
```

# Appendix 4:
# CreatDataSets.m ( Log-return, DFT, Vector projection)

```
clc;
clear all;
target(1,:)={'GOOGL','AAPL','AMZN','BBBY','CSCO','DISCA','DTV',...
    'MSFT','SBUX','YHOO'};%NASDAQ
target(2,:)={'BRBY.L','BT-A.L','BATS.L','IMT.L','IHG.L','ABF.L',...
    'TSCO.L','SBRY.L','MRW.L','EZJ.L'}%FTSE
target(3,:)={'2303.TW','2412.TW','3008.TW','1101.TW','2882.TW',...
    '4938.TW','1303.TW','1301.TW','2002.TW','1216.TW'};%TAIWAN

for tarcount=1:size(target,1)
    
    for i=1:size(target,2)       %get time series
        
[mydate(:,i),myhigh,mylow,myopen(:,i),myclose(:,i),myvol(:,i)]=...
        
get_hist_stock_data(target(tarcount,i),2007,01,04,2009,01,04);
    end
    w=size(myclose,2);        %total number of stock
    n=size(myclose,1);        %total days
    
    %% check data
    for i=1:n
        if sum(strcmp(mydate(i,1),mydate(i,:)))~=10
            error('data not complete');
        end
        
    end
    
    %computing data with 500days
    scale=20;
    segments=25;
    
    %%creat log-returns dataset
```

```matlab
        for jj=1:w

            for i=2:n
                logreturn(i-1,jj)=log(myclose(i,jj)/myclose(i-1,jj));
            end
            for i=1:segments
                dayslogreturn20{i,jj}=logreturn((i-1)*scale+1,jj);
                for j=2:scale
                    dayslogreturn20{i,jj}=[[dayslogreturn20{i,jj}];...
                        logreturn((i-1)*scale+j,jj)];
                end
            end
        end

%create dataset with index
for j=1:w
    for i=1:segments
        if j==1 && i==1
            tempdata=[1,1,dayslogreturn20{1,1}'];
        else
            tempdata=[tempdata;j,i,dayslogreturn20{i,j}'];
        end
    end
end
%create dataset without index
tempdata1=tempdata(:,3:end);

%% Discrete Fourier transform
X=fft(tempdata1',size(tempdata1,2));
X=X';
FS=abs([tempdata(:,1:2),X]);
clear X;

%% projection logreturn compares
C=dayslogreturn20;
```

```matlab
    for i=1:size(C,2)
        for j=1:size(C,2)
            for k=1:size(C,1)
                POJ((i-1)*segments+k,j)= ( (C{k,i}-mean(C{k,i}))' * ...
                    (C{k,j}-mean(C{k,j})) ) / sqrt((C{k,i}-mean(C{k,i}))'...

*(C{k,i}-mean(C{k,i})))*sqrt((C{k,j}-mean(C{k,j}))'...
                    *(C{k,j}-mean(C{k,j})));
            end
        end
    end
    POJ=[tempdata(:,1:2),POJ];
    clear C dayslogreturn20;
    %% copy for market
    LOGmarket{tarcount}=tempdata1;
    FSmarket{tarcount}=FS;
    POJmarket{tarcount}=POJ;
    
    
    clear FS POJ logreturn tempdata tempdata1 mydate myhigh mylow myopen myclose myvol;
    clear i j jj k n w;

end
```

# Appendix 5:
# Function for download the time series table from Yahoo Finance
# Recourse from LuminousLogic.com

I add this code to my thesis for completeness because it is not included into the standard MatLab libraries.

```matlab
% Script to Retrieve Historical Stock Data from Yahoo! Finance
% LuminousLogic.com

function [hist_date, hist_high, hist_low, hist_open, hist_close, hist_vol] = get_hist_stock_data(stock_symbol,beg_year,beg_month,beg_day,fin_year,fin_month,fin_day)

% Define starting year (the further back in time, the longer it takes to download)
%start_year = '2012';

% Get current date
%[this_year, this_month, this_day, dummy, dummy] = datevec(date);

% Build URL string
url_string = 'http://ichart.finance.yahoo.com/table.csv?';
%url_string = 'http://table.finance.yahoo.com/table.csv?';
url_string = strcat(url_string, '&s=', upper(stock_symbol)   );
url_string = strcat(url_string, '&a=', num2str(beg_month-1) );
url_string = strcat(url_string, '&b=', num2str(beg_day) );
url_string = strcat(url_string, '&c=', num2str(beg_year) );
url_string = strcat(url_string, '&d=', num2str(fin_month-1) );
url_string = strcat(url_string, '&e=', num2str(fin_day)    );
url_string = strcat(url_string, '&f=', num2str(fin_year)   );
```

```matlab
url_string = strcat(url_string, '&g=d&a=0&b=1&c=', num2str(beg_year));
url_string = strcat(url_string, '&ignore.csv');

% Open a connection to the URL and retrieve data into a buffer
buffer     = java.io.BufferedReader(...
             java.io.InputStreamReader(...
             openStream(...
             java.net.URL(url_string))));

% Read the first line (a header) and discard
dummy   = readLine(buffer);

% Read all remaining lines in buffer
ptr = 1;
while 1
   % Read line
   buff_line = char(readLine(buffer));

   % Break if this is the end
   if length(buff_line)<3, break; end

   % Find comma delimiter locations
   commas   = find(buff_line== ',');

   % Extract high, low, open, close, etc. from string
   DATEvar   = buff_line(1:commas(1)-1);
   OPENvar   = str2num( buff_line(commas(1)+1:commas(2)-1) );
   HIGHvar   = str2num( buff_line(commas(2)+1:commas(3)-1) );
   LOWvar    = str2num( buff_line(commas(3)+1:commas(4)-1) );
   CLOSEvar  = str2num( buff_line(commas(4)+1:commas(5)-1) );
   VOLvar    = str2num( buff_line(commas(5)+1:commas(6)-1) );
   adj_close = str2num( buff_line(commas(6)+1:end) );
```

```matlab
    %Adjust for dividends, splits, etc.
    DATEtemp{ptr,1} = DATEvar;
    OPENtemp(ptr,1) = OPENvar  * adj_close / CLOSEvar;
    HIGHtemp(ptr,1) = HIGHvar  * adj_close / CLOSEvar;
    LOWtemp (ptr,1) = LOWvar   * adj_close / CLOSEvar;
    CLOSEtemp(ptr,1)= CLOSEvar * adj_close / CLOSEvar;
    VOLtemp(ptr,1)  = VOLvar;
    
    
    ptr = ptr + 1;
end

% Reverse to normal chronological order, so 1st entry is oldest data point
hist_date  = flipud(DATEtemp);
hist_open  = flipud(OPENtemp);
hist_high  = flipud(HIGHtemp);
hist_low   = flipud(LOWtemp);
hist_close = flipud(CLOSEtemp);
hist_vol   = flipud(VOLtemp);
```